\documentclass[acmlarge]{acmart}

\makeatletter
\newcommand{\myconfshort}{\acmConference@shortname}
\newcommand{\myconffull}{\acmConference@name}
\newcommand{\myconfdate}{\acmConference@date}
\newcommand{\myconfloc}{\acmConference@venue}
\AtBeginDocument{
  \fancypagestyle{firstpagestyle}{
    \fancyhead{}%
    \fancyfoot[C]{}%
  }
  \fancyhf{}
  \fancyhead[LO]{\@headfootfont\shorttitle}%
  \fancyhead[RE]{\@headfootfont\@shortauthors}%
  \fancyhead[LE]{\@headfootfont\footnotesize \myconfshort, \myconfdate, \myconfloc}%
  \fancyhead[RO]{\@headfootfont\footnotesize \myconfshort, \myconfdate, \myconfloc}%
  \fancyfoot[C]{}%
}
\makeatother
\acmBooktitle{\conffull\@ (\confshort), \confdate, \confloc}

\AtBeginDocument{%
  }

\usepackage{float}
\usepackage[utf8]{inputenc}
\usepackage{tabularx}
\usepackage{booktabs}

\usepackage{xcolor}
\usepackage{multirow}
\usepackage{longtable}
\usepackage{array}
\usepackage{subcaption}
\usepackage{caption}
\newif\ifshowcomments

\showcommentstrue
\ifshowcomments
  \newcommand{\commentyp}[1]{{\color{blue}[{\bf Yulu:} #1]}}
  \newcommand{\notejs}[1]{{\color{orange}[{\bf Jat:} #1]}}
  \newcommand{\commentll}[1]{{\color{red}[{\bf Lucas:} #1]}}
  \newcommand{\commentjw}[1]{{\color{green}[{\bf Jae Woo:} #1]}}
  \newcommand{\sx}[1]{{\color{magenta}[{\bf Sijia:} #1]}}
    \newcommand{\todo}[1]{{\color{red}[{\bf Todo:} #1]}}
     \newcommand{\noterz}[1]{{\color{violet}[{\bf Renwen:} #1]}}

\else
  \newcommand{\commentyp}[1]{}
  \newcommand{\notejs}[1]{}
  \newcommand{\commentll}[1]{}
  \newcommand{\commentjw}[1]{}
  \newcommand{\sx}[1]{}
    
\newcommand{\todo}[1]{}
 \newcommand{\noterz}[1]{}
\fi

\newcommand{\changeyp}[1]{{\color{black}#1}}

\usepackage{xcolor}

\begin{document}

\copyrightyear{2026}
\acmYear{2026}
\setcopyright{cc}
\setcctype{by}
\acmConference[FAccT '26]{The 2026 ACM Conference on Fairness, Accountability, and Transparency}{June 25--28, 2026}{Montreal, QC, Canada}
\acmBooktitle{The 2026 ACM Conference on Fairness, Accountability, and Transparency (FAccT '26), June 25--28, 2026, Montreal, QC, Canada}
\acmDOI{10.1145/3805689.3812323}
\acmISBN{979-8-4007-2596-8/2026/06}

\title{Push and Pushback in Contesting AI: Demands for and Resistance to Accountability}


\author{Yulu Pi}
\affiliation{%
  \institution{Research Centre Trust, UA Ruhr, University of Duisburg-Essen}
  \city{Duisburg}
  \country{Germany}}
\email{yulu.pi@uni-due.de}

\author{Lucas Lichner}
\affiliation{%
  \institution{Research Centre Trust, UA Ruhr, University of Duisburg-Essen}
  \city{Duisburg}
  \country{Germany}}
\email{lucas@lichner.net}

\author{Jae Woo Lee}
\affiliation{%
  \institution{University of Cambridge}
  \city{Cambridge}
  \country{United Kingdom}}
\email{jwl57@cam.ac.uk}

\author{Sijia Xiao}
\affiliation{%
  \institution{Carnegie Mellon University}
  \city{Pittsburgh}
  \state{Pennsylvania}
  \country{USA}}
\email{sijiaxia@andrew.cmu.edu}

\author{Renwen Zhang}
\affiliation{%
  \institution{Nanyang Technological University}
  \city{Singapore}
  \country{Singapore}}
\email{renwen.zhang@ntu.edu.sg}

\author{Jatinder Singh}
\affiliation{%
  \institution{Research Centre Trust, UA Ruhr, University of Duisburg-Essen}
  \city{Duisburg}
  \country{Germany}}
\affiliation{%
  \institution{University of Cambridge}
  \city{Cambridge}
  \country{United Kingdom}}
\email{jat@compacctsys.net}

\begin{teaserfigure}
   \centering
   \includegraphics[width=\textwidth]{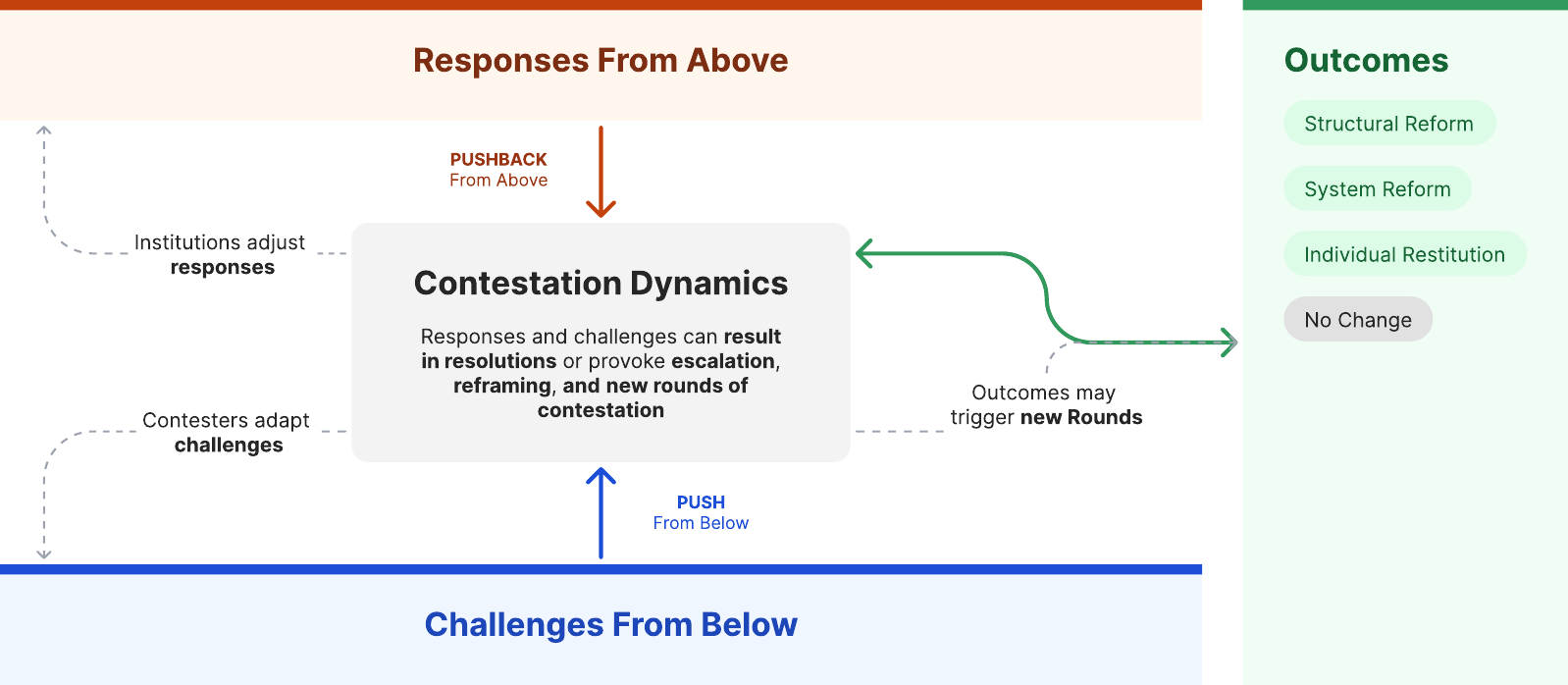}
    \caption{A characterization of AI contestation dynamics, capturing the two-sided accountability relationship between actors from below and actors from above. Contestation emerged as an iterative, prolonged process in which initial challenges provoke institutional responses, which in turn generate new waves of contestation, reframing, and escalation.
    }
    \Description{Description for accessibility.}
    \label{fig:teaser-figure}
\end{teaserfigure}

\begin{abstract}

As AI becomes increasingly embedded in daily life, it has been shown to fail critically, cause harm, and spark public controversy, prompting affected communities, workers, and public-interest groups to contest it. Yet how these contestations unfold in practice remains underexplored. We address this gap by developing an empirically grounded account of AI contestation dynamics. We do so through a thematic analysis of 43 real-world cases in which affected actors direct demands toward those responsible for AI development and deployment, seeking redress, influence, or changes to AI practices. Situating our work within Bovens's relational model of accountability, we conceptualize contestation as accountability-seeking: a dynamic, iterative process in which actors ``from below" direct explicit demands at actors ``from above," who respond by accepting, resisting, or circumventing accountability. Our analysis produces empirically grounded categories of contestation strategies, institutional response tactics, outcome types, and the contextual factors that shape them, illuminating how accountability is pursued and evaded in practice. We show that those being contested often deploy a range of strategies to limit their accountability. Based on these insights, we offer guidance for researchers, policymakers, advocates, and other stakeholders seeking to support effective AI contestation, with particular attention to anticipating and countering institutional strategies used to evade accountability.




\end{abstract}

\maketitle

\section{Introduction}

As AI systems become increasingly embedded in everyday life, it has been shown that they can fail in critical ways, inflict harm, and provoke public controversy. Prior work documents these harms and examines how accountability for AI can be realized~\cite{DarkCompanionship,SociotechnicalHarms,FallAlgorithm,ComesAfterHarm,bottomAlgorithmic}. Much of this research has focused on actions \textit{``from above''}: how actors that design, deploy, or govern AI systems handle accountability and respond to harms~\cite{FallAlgorithm,IncidentstoInsights,RecourseRepair}. At the same time, a growing body of research examines other forms of accountability-seeking, often less `adversarial', such as participation~\cite{StakeholderParticipation, StakeholderInvolvement}, co-design~\cite{Co-DesigningChecklists}, and user feedback mechanisms~\cite{googlepair_feedback_controls}, which rely on collaboration or input rather than direct challenge. Although these approaches can help prevent or mitigate certain harms, they do not address all failures. Moreover, such mechanisms are not always meaningfully accessible to affected communities in practice~\cite{groves2023goingpublicrolepublic, StakeholderInvolvement}.





When routes to redress or accountability that rely on collaboration, consultation, and participation fail or are unavailable, affected communities increasingly turn to more adversarial means. Some may practice algorithmic resistance, tactically attempting to circumvent or subvert AI systems and their effects in everyday practice~\cite{Resistanceandrefusal,bonini_trere_2024_algorithms_of_resistance,AlgorithmAuditing}. Yet beyond tactical resistance, communities ``from below'' also actively contest AI systems and the socio-technical arrangements behind them through complaints, legal actions, media engagement, and public campaigns, explicitly demanding change and accountability from those responsible for their design, deployment, and governance~\cite{aiid:96:houston,aiid:639,bbcnews2020alevels}.


\textit{Contestation} arises when affected individuals and communities challenge decisions and actively seek justification, remedies, or change. In doing so, they compel those in positions of power to respond. This makes contestation an important mechanism through which accountability is pursued. We therefore conceptualize contestation as a distinct modality of accountability-seeking: it is not merely a managed procedure but a relational and interactive process between differently positioned actors. Yet, this two-sided dynamic---the interplay between challenge \textit{from below} and response \textit{from above}---remains largely underexplored. 




The language of ``above'' and ``below,'' with roots in science and technology studies~\cite{harding2008sciences} and applied to algorithmic systems by Whitney et al.~\cite{HCITactics} and Johnson et al.~\cite{FallAlgorithm}, helps characterize the relative positions and power dynamics of actors in relation to these systems. Actors \textit{``from above''} typically occupy positions of power, such as companies that develop AI systems, organizations that deploy them, or authorities that govern them, while actors \textit{``from below''} include affected communities, workers, civil society groups, and concerned publics~\cite{FallAlgorithm}. Across contexts, actors from below make demands on actors above, seeking to influence, redirect, or change specific AI decisions, AI systems as a whole, or the conditions that shape them, prompting responses from above that range from acceptance to resistance. This demand-and-response relationship can be further understood through Bovens's theory of accountability~\cite{BovensAccountability,Bovens01092010,Whattoaccount}: actors from below act as \textit{accountability forums}, questioning and challenging the decisions, practices, and conditions surrounding AI systems, while actors above function as \textit{accountability actors}, responding by accepting, resisting, or circumventing these demands. Our analysis below further demonstrates that this demand-and-response relationship is not a one-off event, but an interactive, dynamic, and often prolonged process. 






Some high-profile incidents illustrate the range and significance of bottom-up contestations to AI: UK students forced the withdrawal of the A-level grading algorithm~\cite{bbcnews2020alevels,Tieleman_2025}, local activists blocked data centers in Indiana and Missouri~\cite{Diode2024}, and employee protests halted Google's Project Maven contracts~\cite{Negotiatingevil,Simonite2021WiredMaven}. Despite the growing visibility of such contestations, \textbf{we still lack a grounded and in-depth understanding of how these dynamics between those who contest AI and those being contested unfold}. Studying these dynamics is crucial: without understanding how contestations unfold in practice, those challenging AI may struggle strategically to identify effective leverage points, anticipate institutional response, and generally to learn from prior efforts.




%
This paper therefore asks: 
\begin{itemize}
    \item \textbf{RQ1 (Contest From Below):} Who contests which aspects of AI, and in what ways?
    \item \textbf{RQ2 (Responses From Above):} How do those being contested respond, and by what strategies?
    \item \textbf{RQ3 (Outcomes):} What outcomes emerge from these interactions, and what shapes them? 
\end{itemize}


We synthesize answers to these questions into structured categories that capture who contests, what is contested, how institutions respond, what outcomes emerge, and which factors shape these dynamics, thereby providing a grounded, in-depth account of how AI contestation unfolds in practice. Unlike prior work focusing on algorithmic resistance and abandonment (see \S\ref{sec:resistance}), \textit{our analysis considers the full relational dynamics between actors from below and above}, indicating how contestation arises, how institutions respond, and how these interactions evolve over time. By centering both sides of this dynamic, this study advances empirically grounded categories of contestation types, institutional response strategies, and outcome patterns, together with the contextual factors that shape why some efforts succeed while others fail. Our analysis also shows that AI contestation is rarely settled cleanly; rather, it is an ongoing negotiation between actors from below and above, with success that is often partial, reversible, or limited in scope. \changeyp{Across cases, we identify several recurring factors that shape contestation outcomes, including the attribution and tangibility of harm; coalition formation and public visibility; institutional cost-benefit calculations and the availability of viable alternatives to the contested system or practice; enforcement mechanisms and jurisdictional capacity, as well as problem definition and strategic framing. Importantly, these factors do not operate independently; instead, they interact and shift in significance as contestation unfolds, requiring alignment across multiple factors to support contestation effort.}


\changeyp{Based on these findings, we offer \textit{concrete guidance for those seeking to contest AI systems} and for those \textit{designing mechanisms to support} such efforts (see \S\ref{lessons}). These include documenting harms and mapping supply chains early in the contestation process; framing challenges in terms of procedural rights rather than solely as technical disputes; treating implementation changes as a distinct stage that requires sustained pressure beyond an initial victory; and investing in shared tools and infrastructures that connect isolated efforts into durable collective capacity across cases. In doing so, we contribute to empowering those contesting AI with the knowledge and strategies needed to demand, negotiate, and eventually hold institutions to account.}

\section{Background and Related Work}



\subsection{Algorithmic Resistance and Algorithmic Contestation}
\label{sec:resistance}

One common way individuals and communities from below sometimes push back against algorithmic systems is through \textit{algorithmic resistance} -- inventive, often tactical strategies that reactively subvert, evade, or mitigate algorithmic power~\cite{Velkova2019AlgorithmicRM,ObfuscationBrunton,Non-Use,AlgorithmAuditing}. However, resistance from below operates to navigate or circumvent the system and the power around it, and therefore does not typically establish a meaningful accountability relationship with actors from above. In contrast, our focus here is on algorithmic contestation, which foregrounds explicit demands directed at identifiable actors and institutions. Importantly, algorithmic contestation directly engages the accountability relationship. Actors \textit{from below} assert their standing as stakeholders and call actors \textit{from above} to account. This might occur through demanding explanations, justifications, remedies or compensation, policy change, or structural reform. The action and ability to contest, sometimes referred to as \textit{contestability}~\cite{contestcon, EmpoweringIndividuals}, provides a form of protection, empowering individuals to regain some control and influence decision-making~\cite{contestcon, kaminski2021, democraticcontrol}. In this paper, we examine both sides of this contestation dynamic: how contestation emerges \textit{from below} and how it is met \textit{from above}, with the goal of understanding the strategies, interactions, and conditions that shape accountability in practice.

\subsection{Accountability as Relational, Contestation as Accountability-Seeking}

Bovens's model~\cite{Bovens01092010,BovensAccountability}, widely drawn on in the algorithmic accountability literature~\cite{Reviewable, Understandingaccountability, Whattoaccount}, conceptualizes accountability as an inherently relational process: an actor \textit{A} is obliged to explain and justify its conduct to a forum \textit{F}, which then may respond, e.g., by posing questions, evaluating the justifications offered, passing judgment, and 
imposing consequences~\cite{whatisAccountability,Bovens01092010,BovensAccountability,Understandingaccountability}. 

Building on this relational understanding, we characterise contestation by actors from below as ``accountability-seeking'': they actively demand explanations, remedies, or institutional changes from those who design, deploy, or govern AI systems. This characterisation has two implications for our analysis. First, it clarifies that there is a specific need to focus on contestation, rather than on the full range of accountability mechanisms such as co-design, participation, or user feedback. Contestation involves explicit demands directed at identifiable actors and a clear assertion of forum standing, making it a distinct and particularly visible modality of active accountability-seeking. Second, it draws attention to the contingent nature of accountability~\cite{Socialaccountability}: actors from below may or may not prompt change or a response, while actors from above may comply, negotiate partial change, deflect responsibility, or refuse accountability altogether. Our analysis therefore explicitly traces both sides of this dynamic: how actors from below enact accountability-seeking through contestation and how actors from above respond in ways that enable or obstruct the realization of accountability.

\section{Methodology}

To capture the full contestation dynamic, we adopt a qualitative case study approach. This method, common in prior research on documented AI incidents (e.g.~\cite{FallAlgorithm, AlgorithmAuditing}), is well suited to the iterative and context-dependent nature of contestation processes. Rather than prioritizing scale or exhaustive coverage of all AI contestation, we focused on analytical depth and diversity, collecting cases until thematic saturation was reached. The selected cases span different forms and targets of contestation, response strategies, and algorithmic domains (e.g., health, education, insurance, policing)

We identified relevant instances through a combination of academic literature, media coverage, and the AI incident database (AIID).\footnote{The AI Incident Database is a publicly accessible collection of documented real-world incidents involving AI (\url{https://incidentdatabase.ai/}).} This approach enables us to select well-documented, high-impact real-world cases that, while not exhaustive of all contestation, are contextually rich and empirically grounded, providing the variation and depth needed to identify key dynamics in AI contestation.

\begin{itemize}

     \item AIID: We conducted keyword searches using terms such as ``protest,'' ``petition,'' ``lawsuit,'' ``sue,'' ``strike,'' ``accuse,'' ``call for,'' and ``allege'' to identify cases involving contestation. These keywords were deliberately diversified to avoid over-representing litigation-based cases~\cite{ainow_litigating_algorithms_2018} and to capture multiple forms of contestations. The search returned 296 results. We retained only those cases whose titles and brief descriptions contained at least one of the specified search terms and met our inclusion criteria. 
 
    \item Academic literature: We drew primarily on Johnson et al.~\cite{FallAlgorithm} (40 cases) and Richards et al.~\cite{IncidentstoInsights} (48 cases), two systematic studies examining AI harm incidents and institutional decisions to discontinue algorithmic systems.
    
    \item Media and news coverage: We included media reports to capture additional high-profile or widely reported instances of contestation that were not covered in the academic literature or database sources. 
\end{itemize}

After removing duplicates across sources, we identified 87 candidate cases, which we then screened using our \textit{inclusion and exclusion criteria}. We included cases in which identifiable actors from below directed contestation toward identifiable actors from above, taking concrete actions that articulated specific demands. We excluded cases limited to rhetorical interventions, such as manifestos or scholarly critiques~\cite{cifor_feminist_data_manifest_no_2019}, as well as general online discussions, on the grounds that these lacked actionable demands or direct engagement with specific actors. This criterion ensured that all selected cases involved direct interaction between those contesting AI and those being contested. Case collection and initial coding proceeded in parallel and continued until we achieved thematic saturation, that is, until additional cases no longer introduced new instances across our core analytical dimensions, consistent with qualitative research standards~\cite{Saturation}. This process resulted in a final dataset of 43 cases.


We conducted a thematic analysis~\cite{IterativeThematicInquiry} of the 43 cases, drawing on multiple types of evidence for each case: media reports, official statements by individuals and organizations, regulatory findings, and legal documents. Thematic analysis is a qualitative method for identifying, analyzing, and reporting patterns within data~\cite{IterativeThematicInquiry}. Three researchers coded the cases iteratively (scheme in Appendix~\ref{codingschema}), combining deductive coding on predefined dimensions (actors, targets, forms of contestation, and outcomes) with inductive coding to capture emergent patterns~\cite{bingham2023,chandra2019}. For example, tracing full contestation timelines emerged inductively, revealing their prolonged and adaptive nature.

To ensure coding consistency, we implemented a structured, sequential validation process in which all coding decisions were subject to cross-author review. First, one author conducted primary coding across all cases and developed detailed case timelines. To ensure independent review, the two other authors each coded half of the cases, comparing their results against the primary coder’s work. Disagreements arose in 12 cases, primarily due to evidentiary gaps in the initial coding, where the reviewing author identified additional sources that provided a more complete picture of the contestation than initially captured. These disagreements were resolved after the reviewing author gathered the additional evidence, including supplementary news reports, updated regulatory decisions, or court documents, and all three authors collectively discussed the cases to reach consensus on the established facts through iterative discussion. Finally, the original coder incorporated the agreed revisions into the case records. This validation process ensured systematic cross-checking and consistent application of the coding scheme across all cases.

\section{Results}

\begin{figure}[h]
    \centering
    \includegraphics[width=1\linewidth]{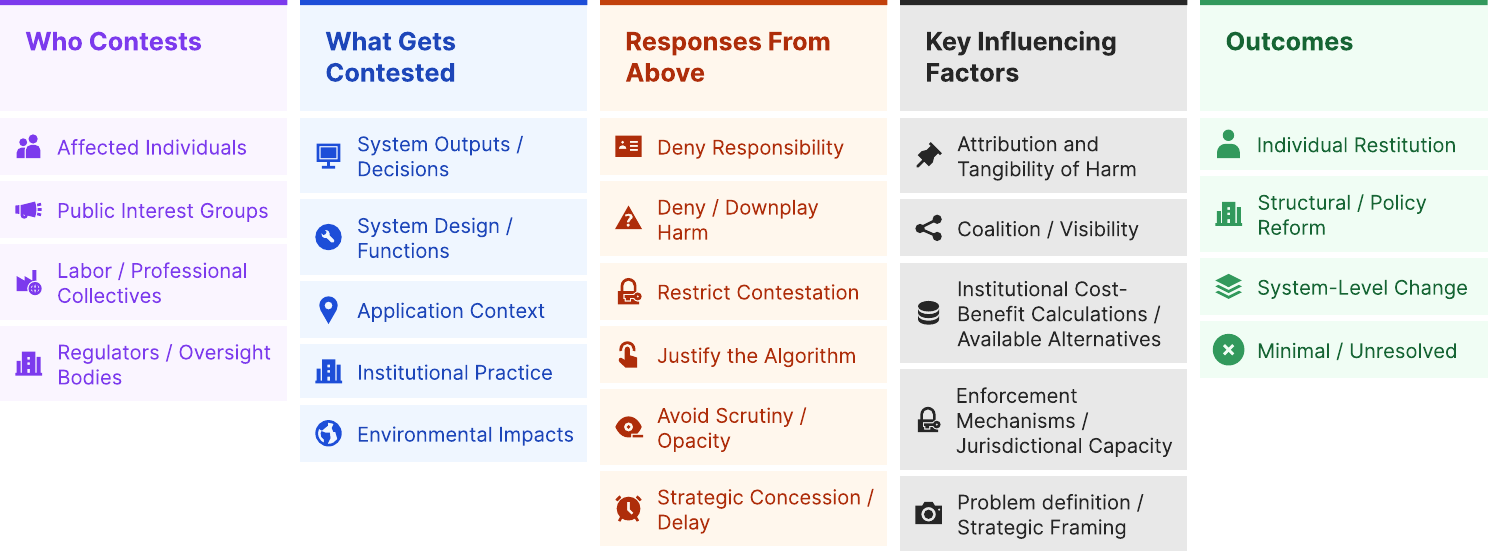}
    



    \caption{We identify four types of contesters, five dimensions of AI they contest, and six institutional response strategies. Five contextual factors mediate these interactions and account for why contestation produces divergent outcomes across individual, system, and structural levels. }
    
    
    \label{fig:overall}
\end{figure}

Our analysis of 43 cases (See Table \ref{tab:overalldata} and Appendix \ref{CaseSelection} for more detailed descriptions) surfaces the specific actors, strategies, and contextual factors that shape how AI contestation unfolds in practice and influence its outcomes (See Figure \ref{fig:overall}). These findings provide a grounded account of relational theories of accountability, offering concrete evidence of how accountability is negotiated in practice. Throughout this section, we reference specific case IDs in the form \textit{C[number]} as \textit{indicative examples} to illustrate and support the points made. 

\subsection{Contest From Below (RQ1)} 
\label{sec:Below}

Our thematic analysis reveals that contestation from below can vary along three intertwined dimensions: \textit{who} initiates the contestation, \textit{what} is being contested, and \textit{how} the contestation is carried out. In practice, contestation can take multiple forms, including direct appeals to organizations (C17, C30, C42) or regulators (C15, C20, C23, C24), public protests (C13, C36, C41), and litigation (C1, C3, C5, C6, C10).


\subsubsection{Who Contests and How They Contest} Patterns of contestation from below vary depending on actors' motivations, capacities, and access to resources. We identify four broad categories of actors:

\textbf{Affected Individuals}
When AI systems cause tangible harm to identifiable individuals, contestation is usually initiated by those directly affected. They pursue available remedies, such as filing complaints (C26, C30) or bringing lawsuits (C6, C22), \textit{in reactive response to} immediate personal loss or harm. Such challenges have targeted misleading or harmful chatbot outputs (C26, C30, C31), copyright infringement (C34), algorithmic allocation decisions (C1, C5, C11), and unfair educational/professional assessments (C10, C13). 

\textbf{ Public-Interest Groups}
These groups can act \textit{in support of affected individuals} (C1, C4, C5) or \textit{on behalf of populations} (C7, C24, C38) exposed to harms that are diffuse, systemic, difficult to attribute to a single individual or group, or legally ambiguous~\cite{ComesAfterHarm,smuha2021beyond,cheong_2024}. Their strategies include \textit{coordinating related cases }(C27), \textit{filing class-action lawsuits} (C3), \textit{engaging regulators} (C23, C38), and \textit{running advocacy or media campaigns} (C21). By mobilizing societal resources, they amplify visibility and access to impacts beyond what affected individuals could achieve alone.

\textbf{Professional, Labor, and Employee Collectives}
Professional associations, labor unions, and employee groups often contest AI deployments even without immediate harms, \textit{seeking to uphold labor standards, industry norms, and ethics} (C36, C37, C39, C40). Contestation may be anticipatory, including \textit{collective action, boycotts, and internal dissent}, as in the 2021 Google Project Nimbus Boycott (C40). These actors draw on their professional expertise, institutional leverage, and networked positions to strengthen the impact of their contestation.

\textbf{Regulators and Oversight Bodies}
Although typically ``from above'', regulators and oversight bodies can act ``from below'' when legal gaps, technological opacity, or cross-jurisdictional barriers limit enforcement. Their strategies include \textit{formal investigations} (C19), \textit{enforcement actions} (C16, C32, C35), or \textit{supporting other actors' contestation} (C2, C15). However, technological complexity, cross-jurisdictional barriers, and legal ambiguities can limit their authority, for example, Clearview AI's noncompliance with GDPR fines due to the absence of EU establishment or assets (C35).

\subsubsection{What Gets Contested} Prior work emphasizes that understanding AI contestation requires contesting beyond harmful system outputs or decisions, adopting a ``supply-chain'' perspective that considers the broader socio-technical arrangements enabling such harms~\cite{Contestability}. Our findings support and extend this claim by providing a fine-grained view of how this broadening unfolds in practice, showing how contestation can both target specific points and move across multiple levels of the broader socio-technical arrangements surrounding AI. Across cases, contestation spans diverse points within this system, from infrastructure and application contexts to institutional practices, system design, and final outputs. These focal points often overlap: contestation may begin with a concrete harm, such as an unfair decision, but typically expands to address underlying causes, system design choices, and organizational practices (C1, C5).



\textbf{System Outputs and Decisions}
A common target is the system's outputs or decisions, such as misleading chatbot responses (C26) or harmful generated content (C27, C30, C31), discriminatory hiring outcomes (C15), and facial-recognition errors leading to wrongful arrests (C25). These harms often serve as entry points for broader contestation of underlying design, deployment, or governance practices.


\textbf{System Design and Functions}
Beyond outputs, contestation also addresses the underlying logic and adequacy of system design. At a narrower level, actors may challenge specific decision rules or variables embedded in the system. For example, the Dutch Childcare Benefits algorithm treated dual and foreign nationality as risk factors, embedding discriminatory logic (C2)~\cite{fenger_simonse_dutch_surveillance_welfare_2024}. At a broader level, contestation can question the adequacy of the system's overall problem formulation or the assumptions that underpin it, such as critiques of a teen pregnancy prediction system in Argentina for oversimplifying complex social problems (C12)~\cite{notmyai_plataforma_intervencion_social_2021}.


\textbf{Application Context} 
Another major focus of contestation concerns the contexts in which AI systems are applied and the purposes they are intended to serve. Contestation arises when systems are introduced into high-stakes or socially sensitive domains, such as grade assessment (C13), public surveillance (C18, C21, C23), or military applications (C39, C40). In these cases, challengers argue that certain uses of AI are inappropriate regardless of technical performance.




\textbf{Institutional Practice}
Contestation also targets the institutional practices that shape how AI systems are developed, procured, governed, and overseen, especially when AI systems cause harm and organizational processes fail to prevent, mitigate, or address those harms. These challenges often focus on a lack of transparency on the use of AI (C5, C18), inadequate human oversight (C1, C14), and exclusion of affected communities from decision-making (C2, C5, C41). 

\textbf{Infrastructure and Environmental Impacts}
Finally, contestation extends beyond specific AI applications to the enabling infrastructure itself. Residents mobilized against data center construction supporting AI operations, raising concerns about land use, environmental impact, and lack of public consultation (C41). This reflects broader contestation of AI's material footprint and enabling conditions, rather than algorithmic harms alone.

\subsection{Responses from Above (RQ2)} \label{responsefromabove}
A general trend observed across cases is that actors being contested tend to push back on contestation through a variety of strategies. A summary of these strategies can be found in Table \ref{tab:responses_from_above_hierarchical} and Figure \ref{fig:overall}. Because our analysis relies on publicly available sources, naturally, there is some focus on contestations that attracted media, regulatory, or other publicly visible attention. 
This does not mean that less visible contestations are unimportant; rather, focusing on observable contestation provides a useful and accessible starting point for unpacking contestation dynamics. Consequently, this section documents 
observable or otherwise visible response strategies, rather than all possible forms of response, such as internal measures taken by actors to address contested issues. We discuss this limitation and its implications in \S\ref{limitations}.

\subsubsection{Denying Responsibility: ``It's Not Our Fault''}
\label{DenyingResponsibility}

When AI systems cause harm, contested actors often deny responsibility, rejecting recognition as parties whose actions or inactions contributed to the resulting harm~\cite{toCourts,DecisionProvenance}. Establishing accountability is inherently challenging because algorithmic systems are technically and organizationally complex, involving multiple actors across AI supply chains. This complexity can make it legally and technically difficult to trace specific harms to a single entity~\cite{Understandingaccountability,Pi_Proctor_2025,COBBE2021105573}, reflecting what Nissenbaum calls the Problem of `Many Hands'~\cite{Accountabilityinacomputerizedsociety}. Moreover, actors can further exploit this ambiguity to distance themselves from responsibility. Two common strategies for denying responsibility emerge from our analysis:

\textit{Shifting responsibility to other actors.~~}
Actors may frame themselves as providers of tools rather than decision-makers, suggesting that ultimate responsibility lies elsewhere. For example, in a collective-action lawsuit alleging that Workday's AI-based applicant-screening tools discriminated against job seekers based on race, age, and disability, Workday argued that its system only provided software to employers and did not make hiring decisions, and thus was not subject to employment anti-discrimination law (C15). By positioning itself as a technical vendor rather than an employer, Workday argued that responsibility for discrimination only rested with its clients, a claim the court ultimately rejected~\cite{Tyman2024MobleyWorkday}.

\textit{Blaming the AI system.~~}
Some actors portray AI as an autonomous agent accountable for its own outputs, shifting responsibility from themselves to the AI system. Air Canada, for instance, responded to complaints about misleading chatbot outputs by claiming that the chatbot ``is a separate legal entity responsible for its own actions''~\cite{aiid:639} (C26). Similarly, Character.AI has claimed that its chatbot outputs should be considered protected speech rather than a product, treating the AI as if it were a human speaker and using this argument to limit the company's liability~\cite{Barcott2025CharacterAI} (C27).




\subsubsection{Denying or Downplaying Harm: ``Nothing serious happened''.~~}
\label{Denyingharms}
When confronted with contestation over algorithmic harms, contested actors may challenge the existence, severity, or systemic nature of harm. Two recurring strategies emerge from our analysis:

\textit{Direct Denial.~}
Actors may reject harm claims entirely or question the credibility of those raising them. For example, NEDA's VP accused a user of fabricating complaints about the organization's Wellness chatbot, only acknowledging the issue after mounting public scrutiny and confirmatory evidence~\cite{FallAlgorithm,Tessa2023Vice} (C30). Similarly, the NYPD initially denied collaborating with Clearview AI, with a spokesperson asserting, ``There is no institutional relationship between the NYPD and Clearview,'' a claim later contradicted by investigative reporting~\cite{hill2020clearview}(C18).

\textit{Downplaying the severity, scope, or scale of harm.~}
Rather than denying harm outright, actors may acknowledge some adverse outcomes but frame them as isolated, minor, or exceptional. This strategy narrows the perceived impact of harm, reducing pressure for systemic intervention. In the Arkansas RUGS case (C1), state officials highlighted that some Medicaid recipients experienced increases in home-care hours~\cite{Hardy2018ARChoicesRuleBlocked}, even as nearly half of recipients experienced substantial cuts~\cite{BTAH_ArkansasHCBSCuts,FallAlgorithm}, creating the impression that harms were limited. Cruise, an autonomous vehicle company, initially omitted key details about dragging a pedestrian when reporting an incident to California regulators, revealing the full severity only after external scrutiny~\cite{aiid:726}(C32). In the Westfield High School deepfake incident (C33), school officials suggested that the images had been deleted and were no longer circulating, framing the harm as resolved and contained despite ongoing reputational and emotional harms for affected students.

Across these cases, denial and downplaying serve as delaying mechanisms, shifting the social, temporal, and financial costs of accountability onto those contesting from below. Even when these strategies do not ultimately prevent findings of harm, as in the cases mentioned above, they force claimants and regulators to expend significant effort validating harm, disproportionately burdening marginalized individuals and other parties with limited resources.

\subsubsection{Restricting Avenues for Contestation: ``You Can't Challenge Us''.~~} 
\label{Restrictingavenues}


Contested actors may also respond by restricting who can challenge them, when challenges can be brought, and the forums in which they can be raised. This approach operates at a fundamental level, using legal, procedural, or institutional mechanisms to narrow or even foreclose avenues for contestation. For instance, Clearview AI faced a lawsuit under the Illinois Biometric Information Privacy Act (BIPA). The company argued that BIPA did not apply to it because it is based in New York, an argument that was ultimately rejected by the court \cite{aclu_clearview_2022} (C18). Contestation can be preemptively limited through contractual design: in Project Nimbus (C40), Google's contract with the Israeli government prevents it from withdrawing services, rendering employee protests ineffective despite their role in ending Project Maven~\cite{solomon_israel_2021, Google2018Maven}.

In the public sector, we also see contested authorities invoke legal exemptions or procedural shields. For example, in the São Paulo Metro case (on appeal) (C24), the city argued that its facial recognition system was exempt from the Brazilian Data Protection Law under Article 4~\cite{Belli_Gaspar_Zingales_2024}. Similarly, the state of Hesse justified the deployment of HessenDATA under the Hessian Act on Public Security and Order (C20); although the Federal Constitutional Court later struck down these provisions as unconstitutional~\cite{Palantirtrust,BVGHessendata2023}, both cases illustrate how legal frameworks can be used to limit judicial scrutiny. Procedural defenses are also visible: in Arkansas RUGS (C1), the state agency argued that plaintiffs had to exhaust administrative remedies before seeking court review, an attempt ultimately rejected by the Supreme Court~\cite{Ledgerwood2017}.

\subsubsection{Justifying the Algorithm: ``AI is the right choice''}
\label{justifyai}

When algorithmic harms are contested, organizations often defend AI systems as the appropriate or superior choice. They do so through two strategies: emphasizing the system's benefits, such as efficiency, accuracy, or public interest, and deflecting responsibility for harms onto human operators. Both approaches help legitimize continued deployment and reduce pressure for systemic change.

\textit{Emphasizing benefits and positive outcomes.~~}
Actors from above may defend contested algorithmic systems by highlighting their benefits and positive outcomes, often framing them in terms of objectivity, efficiency, or public benefit. For example, the Arkansas RUGS case (C1), ultimately abandoned after lawsuits and public outcry over cuts to essential care services~\cite{FallAlgorithm}, was initially defended as ``much more objective'' than human evaluations, which could be prone to ``favoritism and arbitrary decisions''~\cite{IHPI2018}. Similarly, during the 2020 UK A-level grading controversy (C13), then–Prime Minister Boris Johnson described the system as ``robust'' and ``dependable''~\cite{jerseyeveningpost2020alevel}. Comparable justifications appear across domains: Stanford University, when using an algorithm to distribute COVID-19 vaccines (C11), stated that their intent was ``to develop an ethical and equitable process''~\cite{Chen2020}, while systems like HessenDATA (C20) and São Paulo's Metro Intelligent Security System (C24) are defended on grounds of public safety.

\textit{Blaming human operators and defending the system.~~}
Even when harms are undeniable, organizations may attribute errors or adverse outcomes to human operators rather than to the algorithm itself in order to preserve the system's legitimacy. Such attributions may sometimes be valid. This approach both defends the system and limits pressure for systemic changes. For example, In the Detroit facial recognition case (C25), a faulty match resulted in the false accusation of an eight-months-pregnant black woman~\cite{Hill2023} (C25). The Detroit police chief attributed the incident not to the technology, but to ``poor investigative work,'' ~\cite{USCCR2024}, emphasizing that the error lay in the officers' handling of the situation rather than in the technology itself. Similarly, UnitedHealth emphasized that its algorithmic tool is only advisory to human decision-makers, shifting responsibility to human decision-making or existing coverage rules~\cite{guardian2025healthinsurersAI} (C14). This aligns with Elish's warning about the \textit{moral crumple zone}~\cite{MoralCrumpleZones}, which describes how responsibility is obscured for the designers who developed the system and the organizations that deployed it, while frontline human operators with limited control over outcomes bear a disproportionate share of the burden.

Together, emphasizing system benefits and attributing errors to human operators allow organizations to defend the legitimacy and value of the AI system as well as maintain operational continuity despite documented harms. 

\subsubsection{Avoiding Scrutiny Through Intentional Opacity: ``You Are Not Allowed to Know''} 
\label{opacity}


When contestation arises, affected parties need access to evidence to substantiate claims of harm and hold responsible actors accountable~\cite{Pi_Proctor_2025}. Our analysis shows that actors deploying AI systems often use intentional opacity to impede scrutiny~\cite{toCourts,Reviewable} and protect commercial interests~\cite{burrell2016,Wexler2018}. This strategy typically takes two main forms. First, using proprietary or trade-secret protections to block access to audits and obscure system logic. For example, in K.W. v. Armstrong (C5), the State of Idaho claimed its Medicaid budgeting formula was a trade secret~\cite{CoglianeseBenDor2021,Valentine2019}. Similarly, Houston educators were initially obstructed from reviewing the EVAAS value-added model due to proprietary secrecy~\cite{aiid:96:houston} (C10), and UnitedHealth sought to limit discovery in a class-action lawsuit, arguing that broad access to the algorithm's details was unnecessary~\cite{beckers2025unitedhealthAI} (C14).

Second, procedural or institutional opacity arises from arrangements such as non-disclosure agreements (NDAs). In the Peculiar v. Diode Ventures case (C41), NDAs required by the company prevented elected representatives from sharing details of data center projects with local residents~\cite{tannyt}. These cases demonstrate that avoiding scrutiny through intentional opacity, whether via proprietary and trade-secret claims or NDAs, shapes the dynamics of contestation by blocking access to system logic or project information, delaying challenges, and limiting the ability to understand errors, biases, or unfair outcomes.

\subsubsection{Strategic Concession and Delay: ``We'll fix it later (maybe)''} 
\label{Concession}

Actors from above often respond to contestation by limiting or delaying corrective actions, using different methods to manage pressure from contestations without fully addressing systemic issues. This can take several forms: providing case-by-case exceptions or adjustments for individual plaintiffs while resisting system-wide audits or algorithm changes (C1); postponing the implementation of court-mandated reforms over extended periods of nearly ten years (C5); appealing regulatory fines while continuing the contested practice during the appeal process (C16); or making only minor adjustments while leaving flawed systems largely in place (C4). By employing these strategies, organizations maintain the status quo, restrict contestation, and delay meaningful accountability and systemic change.




\subsection{Push and Pushback: The Dynamics of AI Contestation}
\label{sec:factors}



The preceding sections mapped the landscape of AI contestation: who challenges which aspects of AI systems and through what means (\S\ref{sec:Below}), and how institutions respond strategically to those challenges (\S\ref{responsefromabove}). Across our analysis, contestation emerged as an iterative, prolonged process in which initial challenges provoke institutional responses, which in turn generate new waves of contestation, reframing, and escalation. To understand what drives and shapes this process, we inductively identified five key factors that recurrently influenced whether contestation gained or lost momentum across the 43 cases, shaping both its unfolding and its outcomes.



\textbf{Attribution and Tangibility of Harm}: Contestation tended to result in some form of remedy for affected individuals when algorithmic harms are well-documented, and externally traceable to identifiable actors or institutional decisions, such as choices around training data (C34) or failures to follow due process (C1, C5, C10). Clear attribution allows affected individuals to direct claims or demands toward specific entities~\cite{Reviewable, OutliningTraceability}. In contrast, diffuse or cumulative harms distributed across multiple actors or along AI value chains appear harder to contest~\cite{Pi_Proctor_2025, Understandingaccountability}. For instance, in C15 (Workday), alleged discriminatory hiring outcomes were spread across thousands of job applicants and many employer clients. Because Workday operated primarily as an intermediary vendor, responsibility was fragmented, making the case remains unresolved. 

\textbf{Coalition and Visibility}: Contestation can be sustained and amplified when affected individuals, advocacy groups, or other stakeholders form coalitions (C1, C2, C15, C20, C41). Prior work suggests that such coalitions amplify visibility and reputational pressure, which can increase institutional costs of inaction~\cite{FallAlgorithm, MakingChange, stoplapdspying, capp-pgh-2025}. For example, the U.S. Take It Down Act emerged from prolonged advocacy by coalitions of survivors of deepfake harassment and public interest groups (C33), illustrating how sustained contestation can translate into meaningful policy changes. At the same time, coalition is not sufficient on its own: even visible and well-organized coalitions can be dismissed when structural incentives for change remain weak (C23, C43).

\textbf{Institutional Cost-Benefit Calculations and Available Alternatives}: Actors from above tend to weigh the costs and feasibility of complying with contestation demands against the costs of resisting or maintaining the status quo, and these calculations can be as decisive as, or even more decisive than, the significance or visibility of the contestation itself. When readily implementable alternatives to the contested system or practice exist, such as in the UK A-Level algorithm case (C13) or Character.AI's introduction of age verification (C27), institutions can make relatively rapid adjustments. Coalitions can shift institutional cost–benefit calculations by amplifying visibility and reputational pressure, but their influence is constrained by institutional incentives and available alternatives. This is evident in São Paulo's Metro Intelligent Security System (C24), defended on public safety grounds and justified by prior investment, with authorities arguing that abandonment would be more costly than maintenance. Google's maintenance of Project Nimbus (C40) similarly reflects a case where contractual obligations and business interests rendered employee protests structurally ineffective. 

\textbf{Enforcement Mechanisms and Jurisdictional Capacity}: Legal and regulatory structures shape contestation outcomes by enabling or constraining enforcement~\cite{Emmanouilidou2025ClearviewSolomon, extraterritorialenforcement}. For example, U.S. copyright law supported the Anthropic settlement (C34), securing compensation and changes to training data practice. Cross-jurisdictional complexity, under-resourced regulators, or legislative ambiguities can limit changes even when violations are recognized: despite multiple EU data protection authorities issuing fines against Clearview AI, the company has largely avoided payment by arguing it falls outside European jurisdiction as a U.S.-based entity with no local representation or direct services for EU clients (C35), and we often see companies specifying governing law or dispute venue in their Terms of Service, effectively channeling contestations into jurisdictions favorable to them~\cite{Buxbaum_2025}.

\textbf{Problem definition and Strategic Framing}: How contestation is framed influences which outcomes are achievable. Efforts focused on specific cases may secure individual-level outcomes, but system-level change often requires connecting localized incidents to broader patterns~\cite{Pi_Proctor_2025, smuha2021beyond}. For example, in the Character.AI teen harms cases (C27), public interest groups coordinated multiple similar lawsuits; although the individual cases remain ongoing, their collective framing drew attention to recurring risks and contributed to structural changes in the company's service. However, institutions may still resist such reframing by downplaying harms or highlighting algorithmic benefits, as discussed in \S\ref{responsefromabove}. 


\subsubsection{How Factors Interact to Shape Contestation Dynamics and Outcomes}

The above factors do not operate independently. They interact and evolve as contestation unfolds, with each institutional response described in \S~\ref{responsefromabove} continually requiring actors from below to adapt. Three notable patterns recur across cases. 

First, when one avenue of contestation is blocked, actors from below often reframe their challenge to open another. Second, the factors are not interchangeable: a deficit in one cannot be offset by a surplus in another, a strong coalition cannot substitute for enforceable legal mechanisms, and clear harm attribution is insufficient if no feasible alternative exists for the institution being contested. Third, which factor matters most shifts over time: harm attribution is decisive early on, while enforcement capacity becomes the key determinant once a legal victory or formal commitment to change has been secured.

C5 (K.W. v. Armstrong) illustrates all three patterns across fourteen years. When Idaho's trade-secret claims blocked direct scrutiny of the algorithm's logic, plaintiffs reframed their challenge around procedural due process, arguing they had been denied any meaningful opportunity to contest the cuts. This reframing, combined with coalition pressure, opened a legal pathway and secured a court-mandated replacement in 2017. Enforcement capacity then became the key determinant of whether that legal victory would translate into actual change. The state had yet to achieve full implementation by the time of writing.


Comparisons across cases further highlight how different configurations of factors produce divergent outcomes. In some instances, enforcement capacity proves decisive. Both C34 (Anthropic Copyright Lawsuit) and C35 (Clearview AI EU Data Misuse) involved well-documented harms and active coalitions, yet they produced sharply different outcomes because one had access to a forum with enforcement authority while the other did not. In other cases, resolution came quickly when several enabling conditions aligned: tangible harm, clear attribution, and viable alternatives. In C11 (Stanford vaccine), C13 (UK A-levels), C30 (NEDA chatbot), and C31 (Lee-Luda chatbot), harms were immediate and clearly linked to specific institutional decisions, while institutions also had low-cost alternatives available. Coalition pressure and media visibility amplified the salience of these harms, helping mobilize broader public pressure. As a result, all four cases were resolved within the same year the contestation emerged.

By contrast, in cases such as C24 (São Paulo's Metro), C40 (Project Nimbus), and C43 (Pause Letter), coalitions and public visibility were present but insufficient to overcome two compounding obstacles. First, the harms were anticipatory rather than immediate, making attribution contested and reducing the sense of urgency. Second, the demands challenged core institutional interests: public security investment in C24, a legally binding contract in C40, and the broader commercial trajectory of the industry in C43. \changeyp{These cases suggest that when contestation challenges core institutional interests and harms have not yet fully materialized, immediate demands for system withdrawal may be less effective. In such situations, contesters may achieve greater progress by first focusing on building an evidentiary foundation and procedural safeguards. For example, Upadhyay et al. (2026) proposed the use of coordinated Freedom of Information Act (FOIA) requests, which allow the public to access government records and documents, as a strategy for building the evidentiary record needed to support future rounds of contestation~\cite{WorkaroundsAIContestation}.}


Taken together, these comparisons suggest that \textit{coalitions, visibility, and strategic framing can amplify pressure, but they are most effective when combined with clear attribution of harm, feasible alternatives, and mechanisms capable of enforcing change}. Without these enabling conditions, even well-resourced challenges can stall, underscoring how the interaction of factors shapes the dynamics and outcome of contestation.

\subsection{Outcomes (RQ3)}
\label{outcome}
The dynamic interplay between challenges from below and responses from above generates a range of outcomes, from substantive change to purely symbolic responses. While existing frameworks on redress outcome~\cite{Pi_Proctor_2025, unrederss2, RecourseRepair} provide a useful starting point, our analysis shows that contestation frequently extends beyond harm redress to target broader sociotechnical arrangements. In practice, outcomes may emerge simultaneously or sequentially across multiple levels, and contestation does not reliably progress linearly from individual outcomes to system or structural outcomes. Perhaps somewhat counterintuitively, broader changes can occur even without remedies for affected individuals. For example, Character.AI restricted minors' chat access following media scrutiny and regulatory pressure, before any individual litigation was resolved (C27). To capture these variations, we classify contestation outcomes into four levels (see Figure~\ref{fig:overall} and Table~\ref{tab:ai_contestation_outcomes} for more): 


\textbf{Individual-level outcomes (restitution, compensation)} occur when affected individuals receive direct relief addressing the harms they experienced. This includes restoring prior states (e.g., reinstated benefits in C1, corrected grades in C13) or financial compensation (C1, C2, C26). Notably, actors from above often separate individual remedies from broader system-level or structural changes, using the strategic concessions strategy discussed in \S\ref{Concession}.

\textbf{System-level outcomes (repair, abandonment)} arise when institutions modify or remove the AI system itself, %
through repair or full abandonment. Examples include algorithm withdrawals (C11, C13), updates or revisions to systems and processes (C4, C32), and shutdowns (C30, C31). However, even when adjustments or system abandonment occur, similar systems can reemerge~\cite{FallAlgorithm, kalluri2023surveillanceaipipeline}, highlighting the difficulty of achieving lasting system-level change.

\textbf{Structural outcomes (policy reform, service or operational reform)} reshape the broader policies, governance structures, or institutional practices that extend beyond any single system or incident. Examples include Germany's Federal Constitutional Court ruling against §25a HSOG (C20), the U.S. Take It Down Act (C33), Character.AI restricting chats for minors (C27), and the Anthropic settlement requiring destruction of pirated datasets (C30).

\textbf{Minimal or Unresolved outcomes} are symbolic, procedural, or superficial, producing little or no substantive change at any of the levels described above. These tend to arise when challenges lack leverage or tangible harm attribution (C22, C42), when demand remains broad (C43), or when enforcement is weak or delayed (C5, C16, C35).


The factors identified in \S\ref{sec:factors} help explain why contestation produces outcomes at different levels. Individual-level outcomes tend to emerge when harms are clearly attributable and remedies can be provided to identifiable parties. System-level outcomes, by which we mean modifications to or withdrawal of the AI system itself, become more achievable when coalitions amplify pressure and readily implementable alternatives exist. Structural outcomes, however, require an even more demanding convergence of conditions: sustained multi-stakeholder pressure, connecting individual cases to reveal broader patterns, coordination across civil society, regulators, and favorable political conditions. \changeyp{Sustained contestation, combined with a strong convergence of these conditions, can produce structural change by creating new enforcement mechanisms where none previously existed. C20 (HessenDATA) and C33 (Take It Down Act) illustrate this dynamic: five years of civil society pressure in C20 led to constitutional reform of the legal basis for state surveillance, while coordinated advocacy in C33 transformed individual harms into federal legislation criminalizing deepfake abuse. These cases show how contestation can expand the institutional and legal terrain on which future challenges to AI systems can be pursued.}


Yet even where these conditions align and contestation succeeds, our analysis points to a further challenge:  \textit{the contingent and reversible nature of outcomes}. Even when initial victories occur, they are often partial or fragile. Arkansas DHS granted backpay but resisted audits or algorithm withdrawal (C1), and abandoned systems can ``reincarnate''~\cite{FallAlgorithm} elsewhere (C23). Legal and regulatory wins may also take years to implement: the 2017 Idaho Department of Health and Welfare settlement (C5) remained incomplete by 2026, while Amazon continued contested practices while appealing fines (C16). These patterns show that sustained pressure on multiple fronts is often needed to achieve durable changes.



 \

\section{Lessons for Future Contestation Efforts}
\label{lessons}

The analysis above yields a set of empirically grounded categories, covering contestation targets, institutional response strategies, outcomes, and the contextual factors shaping success or failure, that together concretely illustrate how accountability is negotiated through dynamic, iterative interactions between those who contest AI and those being contested. As we show, institutions routinely deploy the strategies identified in \S\ref{responsefromabove} to push back against contestation, often slowing or blocking meaningful change. Effective contestation therefore requires anticipating these responses and preparing for sustained engagement rather than a single decisive confrontation. The lessons summarized in Table~\ref{tab:lessons} are intended as actionable guidance for a range of stakeholders, including affected communities and advocates seeking to contest AI more effectively, as well as policymakers and regulators designing mechanisms to support AI contestation. Each lesson is evidence-based and mapped to the institutional strategies documented in \S\ref{responsefromabove} that it counters.



\begin{table*}[ht]
\centering
\footnotesize
\renewcommand{\arraystretch}{1.3}
\setlength{\tabcolsep}{6pt}
\caption{Empirically grounded lessons for contesting AI, mapped to the institutional strategies documented in \S\ref{responsefromabove} they counter.}
\label{tab:lessons}
\begin{tabularx}{\textwidth}{
    >{\bfseries\raggedright\arraybackslash}p{3.2cm}
    >{\raggedright\arraybackslash}X
    >{\raggedright\arraybackslash}p{4.2cm}
}
\toprule
Lesson &
What to do &
Institutional strategy countered \\
\midrule

Build evidence of harm and accountability &
Document algorithmic harms systematically and map the full AI supply chain early to identify responsible actors. &
Denying responsibility (\S\ref{DenyingResponsibility}); Denying harm (\S\ref{Denyingharms})\\
 
\addlinespace
 
Strategically frame the terms of contestation &
Frame challenges around due process and procedural rights, so that public and policy discussions are not shaped by the institution's own technical justifications. &
Justifying the algorithm (\S\ref{justifyai}); Avoiding scrutiny through opacity (\S\ref{opacity}) \\
 
\addlinespace
 
Treat enforcing implementation as a distinct stage  &
Plan beyond the outcome: ongoing monitoring and sustained pressure are needed to prevent implementation from being delayed, narrowed, or diluted over time. &
Strategic concession and delay (\S\ref{Concession}) 
 \\
 
\addlinespace
 
Lower barriers and enable collective action &
Develop practical tools, reporting platforms, auditing tools, complaint templates, and strengthen procedural rights to enable coordinated challenges that pool resources against systemic harms. &
Restricting avenues for contestation (\S\ref{Restrictingavenues}) \\
 
\addlinespace
 
Build and sustain public-interest infrastructure &
Maintain shared harm repositories, reusable legal and technical tools, and communities of practice that preserve knowledge across cases and transform one-off contestations into enduring challenges. &
All \S\ref{responsefromabove} strategies 
\\
 
\bottomrule
\end{tabularx}
\end{table*}

\textbf{Build Evidence of Harm and Accountability.} Our analysis shows that institutional resistance often begins with denying harm (see \S\ref{Denyingharms}) or obscuring responsibility (\S\ref{DenyingResponsibility}). Contestation therefore requires systematically documenting both the impacts of AI systems and the actors involved in producing them. \textit{On the harm side}, for example, some approaches involve creating spaces where affected individuals can report, discuss, and document their experiences with algorithms through reporting platforms, advocacy networks, or community-led documentation efforts~\cite{AlgorithmAuditing,dataharmrecord, ajl_harms,OutliningTraceability,AccountabilityInfrastructure}. Analyzing these reports across cases can help transform seemingly isolated incidents into evidence of systemic harm. 
\textit{On the accountability side}, mapping the full AI supply chain early, identifying the actors involved in design, procurement, and deployment, can help direct demands toward the right parties. Because responsibility is often distributed across multiple actors~\cite{Understandingaccountability, RecourseRepair}, effective contestation may, depending on available resources, need to target several actors simultaneously. 


\textbf{Strategically Frame the Terms of Contestation.} Institutions frequently defend their AI by emphasizing accuracy and efficiency (\S\ref{justifyai}), and may exploit opacity to resist audits or external scrutiny (\S\ref{opacity}). When contestation remains confined to technical debates on these terms, it risks being shaped by the institution's own justifications. Our analysis suggests that broadening the framing of challenges around due process and procedural rights can help open up different avenues for change. For example, Houston teachers (C10) and the plaintiffs in K.W. v. Armstrong (C5) framed their challenges around due process and procedural rights, securing favorable settlements in court. Collaboration with advocacy organizations or journalists early on can help ensure that public and policy discussions are shaped by these broader considerations, rather than being dominated by the institution's technical narrative.


\textbf{Treat Enforcing Implementation as a Distinct Stage of Contestation.} Even when contestation produces a favourable outcome that mandates changes, such as a court ruling, a regulatory decision, or a public commitment to reform, the institutions being contested often delay or selectively implement the required changes (C5, C16). This gap between formal outcome and material change suggests that enforcing implementation should be understood as a distinct and necessary stage of contestation in its own right, rather than an automatic consequence of winning. Accordingly, those engaged in or supporting contestation must anticipate and prepare for this extended phase, which demands ongoing monitoring, follow-up action, and sustained pressure well beyond any initial win.





\textbf{Lower the Barriers to Contestation and Enable Collective Action.} Contestation often requires legal, technical, and organizational resources that many affected individuals or communities lack~\cite{Pi_Proctor_2025}. Lowering these barriers is therefore essential to making contestation both accessible and sustainable. At a structural level, this involves ensuring that the conditions for contestation are in place, particularly by making procedural rights to explanation and redress meaningfully exercisable in practice~\cite{kaminski2021, noyb_exercise_your_rights_2025}. At a practical level, it entails developing and disseminating tools and infrastructures that support contestation, for example, harm-reporting platforms, accessible auditing tools, and standardized complaint or appeal templates~\cite{ResponsesFromBelow, AccountabilityInfrastructure, WorkaroundsAIContestation}. Because AI-related harms, such as discrimination, are inherently collective, enabling and supporting coordinated forms of contestation, like class actions, community-led challenges, or joint advocacy efforts, allows contesting actors to pool resources and challenge systemic harms that would be difficult to address individually. 



\textbf{Build and Sustain Capacity and Infrastructure for Contestation.} Individual contestation efforts, however well-resourced, remain episodic unless supported by durable capacity and infrastructure that connects expertise, evidence, and coordinated action across cases. Shared repositories of documented harms, reusable legal and technical tools, and communities of practice that preserve knowledge and lessons across contestations can begin to address this, not by winning any single case, but by making the broader landscape of AI contestation more navigable over time. How such capacity and infrastructure can be built, sustained, and what forms it needs to take across different contexts remains an important open question this study leaves for future research.

\section{Limitations} 
\label{limitations}

This study relies on publicly available sources and therefore captures forms of contestation visible through media coverage, litigation, or public controversy, more likely reflecting high-profile incidents than quieter, less-reported instances of harm. Importantly, however, our study's purpose was to be exploratory, and by deliberately focusing on publicly observable dynamics and selecting cases iteratively across multiple sources until thematic saturation was reached, we were able to uncover recurring patterns across a wide range of contestations and their broader practical effects. Moreover, publicly visible contestation is itself consequential, providing a meaningful starting point for understanding how contestation dynamics unfold in practice. Future research could extend this work through interviews, ethnography, or other methods to analyse less visible contestation dynamics and the internal logics of institutional response.

\section{Conclusion}

Bottom-up challenges to AI are crucial for exposing algorithmic harms and demanding accountable systems. Drawing on a thematic analysis of 43 real-world contestation cases spanning diverse actor types, algorithmic domains, and contestation dynamics, this paper provides an empirically grounded, two-sided account of AI contestation as an iterative and prolonged negotiation for accountability. Actors from below mobilize, articulate harms and concerns, and persist in keeping them visible, while actors from above respond through a range of strategies that seek to absorb, defer, or deflect responsibility. Change, when it occurs, is neither automatic nor linear; it is negotiated over time shaped by factors such as how harms are defined and attributed, whether alternatives are feasible, and what enforcement mechanisms are available. Understanding these dynamics equips actors contesting AI to be better informed and more strategic in their efforts, while supporting the design of governance interventions that are better aligned with how contestation operates in practice.

\newpage

\begin{acks}
We acknowledge the support of the UKRI Engineering and Physical Sciences Research Council (EPSRC) grant EP/W020548/1.
\end{acks}

\section{Generative AI Usage Statement}
ChatGPT (GPT-5.2) and Gemini 3 were used for limited editorial and technical support, including citation formatting, table structuring and layout adjustments, literature discovery, and grammar and style checking.

\bibliographystyle{ACM-Reference-Format}
\bibliography{reference, cases}

@misc{IdahoACLUCase,
  author       = {ACLU of Idaho},
  title        = {K.W. v. Armstrong},
  year         = {2025},
  note         = {Accessed December 16, 2025}
}

@misc{ArkansasMedicaidHCBSCuts,
  author       = {{Benefits Tech Advocacy Hub}},
  title        = {Arkansas Medicaid Home and Community Based Services Hours Cuts},
  year         = {2022},
  url = {https://www.btah.org/case-study/arkansas-medicaid-home-and-community-based-services-hours-cuts.html},
  note         = {Accessed 2025-12-16}
}

@misc{BTAH_ArkansasHCBSCuts,
  author       = {Benefits Tech Advocacy Hub},
  title        = {Arkansas Medicaid Home and Community Based Services Hours Cuts},
  howpublished = {\url{https://www.btah.org/case-study/arkansas-medicaid-home-and-community-based-services-hours-cuts.html}},
  note         = {Accessed: 2025-12-03}
}

@misc{Ledgerwood2017,
  author       = {{Justia U.S Law}},
  title        = {Arkansas Department of Human Services v. Ledgerwood (Majority)},
  year         = {2017},
  howpublished = {\url{https://law.justia.com/cases/arkansas/supreme-court/2017/cv-17-183.html}},
  note         = {Accessed 2025-12-08}
}

@misc{tannyt,
  author       = {The New York Times},
year = {2024},
  title        = {A rural Missouri town fights Big Tech, and itself}
}

@misc{guardian2025healthinsurersAI,
  author       = {{The Guardian}},
  title        = {New AI tool counters health insurance denials decided by automated algorithms},
  howpublished = {\url{https://www.theguardian.com/us-news/2025/jan/25/health-insurers-ai}},
  year         = {2025},
}

@misc{Tyman2024MobleyWorkday,
  author       = {Annette Tyman},
  title        = {Mobley v. Workday: Court Holds AI Service Providers Could Be Directly Liable for Employment Discrimination Under “Agent” Theory},
  year         = {2024},
  howpublished = {\url{https://www.seyfarth.com/news-insights/mobley-v-workday-court-holds-ai-service-providers-could-be-directly-liable-for-employment-discrimination-under-agent-theory.html}},
}

@misc{futureoflife2023pause,
  title        = {Pause Giant {AI} Experiments: An Open Letter},
  author       = {{Future of Life Institute}},
  year         = {2023},
  month        = mar,
  howpublished = {\url{https://futureoflife.org/open-letter/pause-giant-ai-experiments/}},
  note         = {Accessed 2025-12-16}
}

@misc{AnthropicHungerStrike2025, 
title={Hunger strike against AGI in San Francisco passes one-month mark}, 
url={https://peninsulapress.com/2025/10/02/hunger-strike-against-agi-passes-one-month-mark/}, 
journal={Peninsula Press}, 
author={Eidesvik, Anders}, 
year={2025}, 
month=oct, 
language={en-US},
  note         = {Accessed 2025-12-16}
}

@inproceedings{oh2025leeluda,
  author    = {Oh, Jiwon Jenn},
  title     = {Navigating Gendered Anthropomorphism in {AI} Ethics: The Case of Lee Luda in South Korea},
  booktitle = {Proceedings of the 58th Hawaii International Conference on System Sciences},
  year      = {2025},
  pages     = {6784--6793},
  doi       = {10.24251/HICSS.2025.809},
  url       = {https://hdl.handle.net/10125/109659},
}

@article{Diode2024, 
title={Fighting back against data centers, one small town at a time}, 
url={https://www.washingtonpost.com/technology/2024/10/05/data-center-protest-community-resistance/}, 
journal={The Washington Post}, 
author={O’Donovan, Caroline}, 
year={2024}, 
}

@online{Google2018Maven,
  author       = {Business and Human Rights Centre},
  title        = {Google introduces “Artificial Intelligence principles” that prohibit its use in weapons \& human rights abuses},
  year         = {2018},
  month        = jul # "~18",
  url          = {https://www.business-humanrights.org/en/latest-news/google-introduces-artificial-intelligence-principles-that-prohibit-its-use-in-weapons-human-rights-abuses/},
  organization = {Business and Human Rights Centre},
  note         = {Accessed: 2025-12-19}
}

@misc{swinhoe2021nimbus,
  author       = {Swinhoe, Dan},
  title        = {Israel Government Says {AWS} and {Google} Can’t Boycott {Nimbus} Project},
  howpublished = {Data Center Dynamics},
  year         = {2021},
  month        = may,
  day          = {25},
  url          = {https://www.datacenterdynamics.com/en/news/israel-government-says-aws-and-google-cant-boycott-nimbus/},
  note         = {Accessed 2025-12-16}
}

@online{solomon_israel_2021,
  author       = {Solomon, Shoshanna},
  title        = {Israel signs deal for cloud services with Google, Amazon},
  date         = {2021-05-24},
  url          = {https://www.timesofisrael.com/israel-signs-deal-for-cloud-services-with-google-amazon/},
  organization = {The Times of Israel},
  urldate      = {2024-05-20},
  language     = {en-US}
}

@online{aclu_clearview_2022,
  author        = {{American Civil Liberties Union}},
  title         = {ACLU v. Clearview AI},
  year         = {2022},
  month        = may,
  url          = {https://www.aclu.org/cases/aclu-v-clearview-ai},
  urldate      = {2024-05-20},
  organization = {American Civil Liberties Union},
  language     = {en-US}
}

@article{NYPDTheCity2025, 
title={NYPD bypassed facial recognition ban to ID Pro-Palestinian student protester}, 
url={https://www.thecity.nyc/2025/07/18/nypd-fdny-clearview-ai-ban-columbia-palestinian-protest/}, 
journal={THE CITY - NYC News}, 
author={Maldonado, Samantha}, 
year={2025}, 
month=jul, 
language={en-US},
  note         = {Accessed 2025-12-16}
}

@misc{hill2020clearview,
  author       = {The New York Times},
  title        = {New Jersey Bars Police From Using Clearview Facial Recognition App},
year= {2020},
  howpublished = {\url{https://www.nytimes.com/2020/01/24/technology/clearview-ai-new-jersey.html}},
  note         = {The New York Times, January 24, 2020. Accessed: 2026-01-03},
}

@online{NOYB2025ClearviewCriminalComplaint,
  author       = {{European Center for Digital Rights (noyb)}},
  title        = {Criminal complaint against facial recognition company Clearview AI},
  year         = {2025},
  month        = oct # "~28",
  url          = {https://noyb.eu/en/criminal-complaint-against-facial-recognition-company-clearview-ai},
  organization = {noyb.eu},
  note         = {Accessed: 2025-12-19}
}

@online{Emmanouilidou2025ClearviewSolomon,
  author       = {Lydia Emmanouilidou},
  title        = {How a Shady US AI Company Dodged Fines and Defied Regulators Across Europe},
  year         = {2025},
  month        = apr # "~16",
  url          = {https://wearesolomon.com/mag/format/investigation/clearview-how-a-shady-us-ai-company-dodged-fines-and-defied-regulators-across-europe/},
  organization = {We Are Solomon},
  note         = {Accessed: 2025-12-13}
}

@online{Warburton2020ClearviewCanada,
  author       = {Moira Warburton},
  title        = {Clearview AI stops offering facial recognition technology in Canada},
  year         = {2020},
  month        = jul # "~07",
  url          = {https://www.reuters.com/article/technology/clearview-ai-stops-offering-facial-recognition-technology-in-canada-idUSKBN24808G/},
  organization = {Reuters},
  note         = {Accessed: 2025-12-19}
}

@online{Dean2023OpenAIDataLawsuit,
  author       = {Grace Dean},
  title        = {A lawsuit claims OpenAI stole “massive amounts of personal data,” including medical records and information about children, to train ChatGPT},
  year         = {2023},
  month        = jun # "~29",
  url          = {https://www.businessinsider.com/openai-chatgpt-generative-ai-stole-personal-data-lawsuit-children-medical-2023-6},
  organization = {Business Insider},
  note         = {Accessed: 2025-12-19}
}

@misc{clarkson2023openai_privacy,
  title        = {Class Action Challenges OpenAI on Privacy},
  author       = {{Clarkson Law Firm, P.C.}},
  year         = {2023},
  url          = {https://clarksonlawfirm.com/class-action-challenges-openai-on-privacy/},
  note         = {Accessed: 2025-12-12}
}

@online{BvA2025Reuters,
  author       = {Blake Brittain},
  title        = {US judge preliminarily approves \$1.5 billion Anthropic copyright settlement},
  year         = {2025},
  month        = sep # "~25",
  url          = {https://www.reuters.com/sustainability/boards-policy-regulation/us-judge-approves-15-billion-anthropic-copyright-settlement-with-authors-2025-09-25/},
  organization = {Reuters},
  note         = {Accessed: 2025-12-19}
}

@misc{CNIL2024AmazonFine,
  author       = {{Commission nationale de l'informatique et des libertés (CNIL)}},
  title        = {Employee Monitoring: CNIL Fined Amazon France Logistique €32 Million},
  year         = {2024},
  howpublished = {\emph{CNIL}},
  url          = {https://www.cnil.fr/en/employee-monitoring-cnil-fined-amazon-france-logistique-eu32-million},
  note         = {Accessed: 2025-12-13}
}

@misc{Reuters2024AmazonAppealsFine,
  author       = {Singh,Jaspreet },
  title        = {Amazon appeals \$34.6 mln fine by French regulator over staff monitoring},
  year         = {2024},
  day          = {22},
  howpublished = {\emph{Reuters}},
  url          = {https://www.reuters.com/technology/amazon-appeals-346-mln-fine-by-french-regulator-over-staff-monitoring-2024-03-22/},
  note         = {Accessed: 2025-12-13}
}

@online{Swaine2014McDonaldsWageTheft,
  author       = {Jon Swaine},
  title        = {McDonald’s lawsuits allege wage theft by fast-food giant and franchise owners},
  year         = {2014},
  month        = mar # "~13",
  url          = {https://www.theguardian.com/business/2014/mar/13/mcdonalds-lawsuits-allege-wage-theft-franchise-owners},
  organization = {The Guardian},
  note         = {Accessed: 2025-12-19}
}

@misc{BVGHessendata2023,
  author       = {{Bundesverfassungsgericht, German Federal Constitutional Court}},
  title        = {Judgment of the First Senate of 16 February 2023 - 1 BvR 1547/19 -, paras. 1-178}, 
  year         = {2023},
  month        = {February},
  howpublished = {\url{https://www.bverfg.de/e/rs20230216_1bvr154719en.html}},
  note         = {English translation. Accessed 2025-12-08}
}

@misc{Chen2020,
  author       = {Chen, Caroline},
  title        = {Only Seven of Stanford’s First 5,000 Vaccines Were Designated for Medical Residents},
  year         = {2020},
  month        = {December},
  howpublished = {\url{https://www.propublica.org/article/only-seven-of-stanfords-first-5-000-vaccines-were-designated-for-medical-residents}},
  note         = {Accessed 2025-12-08}
}

@misc{japannews20250606,
  author = {{The Yomiuri Shimbun}},
  title  = {Newspaper Group Calls on AI Companies to Get Permission to Use News Content; Says Unauthorized Training Could Amount to Copyright Infringement},
  year   = {2025},
  url    = {https://japannews.yomiuri.co.jp/society/general-news/20250606-259586/},
  note   = {Accessed: 18 December 2025},
}

@article{AIisSoulless,
author = {Halperin, Brett A. and Rosner, Daniela K.},
title = {“AI is Soulless”: Hollywood Film Workers’ Strike and Emerging Perceptions of Generative Cinema},
year = {2025},
issue_date = {April 2025},
publisher = {Association for Computing Machinery},
address = {New York, NY, USA},
volume = {32},
number = {2},
issn = {1073-0516},
url = {https://doi.org/10.1145/3716135},
doi = {10.1145/3716135},
journal = {ACM Trans. Comput.-Hum. Interact.},
month = apr,
articleno = {19},
numpages = {27}
}

@misc{capp-pgh-2025,
  author = {{Coalition Against Predictive Policing in Pittsburgh}},
  title  = {Coalition Against Predictive Policing in Pittsburgh – Predictive Policing Advocacy and Resources},
  year   = {2025},
  url    = {https://capp-pgh.com/},
  note   = {Advocacy site opposing predictive policing in Pittsburgh; accessed 18 December 2025},
}

@misc{Hill2023,
  author       = {{The New York Times}},
  title        = {Eight Months Pregnant and Arrested After False Facial Recognition Match},
  year         = {2023},
  howpublished = {\url{https://www.nytimes.com/2023/08/06/business/facial-recognition-false-arrest.html}},
  note         = {Accessed 2025-12-08}
}

@misc{Tessa2023Vice,
  author       = {Xiang, Chloe},
  title        = {Eating Disorder Helpline Disables Chatbot for ‘Harmful’ Responses After Firing Human Staff},
  howpublished = {\url{https://www.vice.com/en/article/eating-disorder-helpline-disables-chatbot-for-harmful-responses-after-firing-human-staff/}},
  year         = {2023},
  note         = {Accessed: 2025-12-03}
}

@misc{Barcott2025CharacterAI,
  author       = {Barcott, Bruce},
  title        = {In early ruling, federal judge defines Character.AI chatbot as product, not speech},
  howpublished = {\url{https://www.transparencycoalition.ai/news/important-early-ruling-in-characterai-case-this-chatbot-is-a-product-not-speech}},
  note         = {Accessed: 2025-12-03},
  year         = {2025}
}

@article{Abelson2024AppleCSAM,
    author = {Abelson, Harold and Anderson, Ross and Bellovin, Steven M and Benaloh, Josh and Blaze, Matt and Callas, Jon and Diffie, Whitfield and Landau, Susan and Neumann, Peter G and Rivest, Ronald L and Schiller, Jeffrey I and Schneier, Bruce and Teague, Vanessa and Troncoso, Carmela},
    title = {Bugs in our pockets: the risks of client-side scanning},
    journal = {Journal of Cybersecurity},
    volume = {10},
    number = {1},
    pages = {tyad020},
    year = {2024},
    month = {01},
    issn = {2057-2085},
    doi = {10.1093/cybsec/tyad020},
    url = {https://doi.org/10.1093/cybsec/tyad020},
    eprint = {https://academic.oup.com/cybersecurity/article-pdf/10/1/tyad020/61182335/tyad020.pdf},
}

@online{Wadsworth2024SFPDFacialRecognition,
  author       = {Jennifer Wadsworth},
  title        = {SFPD skirted facial-recognition ban, lawsuit says. Hundreds of cases could be in jeopardy},
  year         = {2024},
  month        = jul # "~18",
  url          = {https://sfstandard.com/2024/07/18/san-francisco-police-facial-recognition-violations/},
  organization = {The San Francisco Standard},
  note         = {Accessed: 2025-12-19}
}

@online{Greschler2024PropESFPDFacialRecognition,
  author       = {Gabe Greschler},
  title        = {Voters approve Prop. E, giving more powers to San Francisco police},
  year         = {2024},
  month        = mar # "~05",
  url          = {https://sfstandard.com/2024/03/05/san-francisco-voters-election-prop-e-results-police/},
  organization = {The San Francisco Standard},
  note         = {Accessed: 2025-12-19}
}

@online{Sheard2019SFPDFacialRecognition,
  author       = {Nathan Sheard},
  title        = {San Francisco takes a historic step forward in the fight for privacy},
  year         = {2019},
  month        = may # "~14",
  url          = {https://www.eff.org/deeplinks/2019/05/san-francisco-takes-historic-step-forward-fight-privacy},
  organization = {Electronic Frontier Foundation},
  note         = {Accessed: 2025-12-19}
}

@online{BvA2025DAIL,
  author       = {{Ethical Tech Initiative, George Washington University}},
  title        = {Bartz v. Anthropic PBC},
  year         = {2024},
  url          = {https://blogs.gwu.edu/law-eti/ai-litigation-database/case-detail-page/?pid=251},
  note         = {DAIL – Database of AI Litigation. Accessed: 2025-12-19}
}

@article {
      aiid:75:googleinstant,
      author = {Anonymous},
      editor = {McGregor, Sean},
      journal = {AI Incident Database},
      publisher = {Responsible AI Collaborative},
      title = {Incident Number 75: Google Instant's Allegedly 'Anti-Semitic' Results Lead To Lawsuit In France},
      url = {https://incidentdatabase.ai/cite/75},
      year = {2012},
      urldate = {December 19, 2025},
      note = {Retrieved December 2025 from 
        \url{https://incidentdatabase.ai/cite/75}}
}

@article {
      aiid:91,
      author = {McGregor, Sean},
      editor = {McGregor, Sean},
      journal = {AI Incident Database},
      publisher = {Responsible AI Collaborative},
      title = {Incident Number 91: Frontline workers protest at Stanford after hospital distributed vaccine to administrators},
      url = {https://incidentdatabase.ai/cite/91},
      year = {2020},
      urldate = {November 20, 2025},
      note = {Retrieved November 2025 from 
        \url{https://incidentdatabase.ai/cite/91}}
}

@article {
      aiid:96:houston,
      author = {Lutz, Roman},
      editor = {McGregor, Sean},
      journal = {AI Incident Database},
      publisher = {Responsible AI Collaborative},
      title = {Incident Number 96: Houston Schools Must Face Teacher Evaluation Lawsuit},
      url = {https://incidentdatabase.ai/cite/96},
      year = {2017},
      urldate = {December 9, 2025},
      note = {Retrieved December 2025 from 
        \url{https://incidentdatabase.ai/cite/96}}
}

@article {
      aiid:184,
      author = {Anonymous},
      editor = {McGregor, Sean},
      journal = {AI Incident Database},
      publisher = {Responsible AI Collaborative},
      title = {Incident Number 184: Facial Recognition Program in São Paulo Metro Stations Suspended for Illegal and Disproportionate Violation of Citizens’ Right to Privacy},
      url = {https://incidentdatabase.ai/cite/184},
      year = {2018},
      urldate = {December 18, 2025},
      note = {Retrieved December 2025 from 
        \url{https://incidentdatabase.ai/cite/184}}
}

@article {
      aiid:360:McDBiometric,
      author = {Anonymous},
      editor = {Lam, Khoa},
      journal = {AI Incident Database},
      publisher = {Responsible AI Collaborative},
      title = {Incident Number 360: McDonald's AI Drive-Thru Allegedly Collected Biometric Customer Data without Consent, Violating BIPA},
      url = {https://incidentdatabase.ai/cite/360},
      year = {2021},
      urldate = {December 19, 2025},
      note = {Retrieved December 2025 from 
        \url{https://incidentdatabase.ai/cite/360}}
}

@article {
      aiid:489,
      author = {McNulty, Luna},
      editor = {Lam, Khoa},
      journal = {AI Incident Database},
      publisher = {Responsible AI Collaborative},
      title = {Incident Number 489: Workday's AI Tools Allegedly Enabled Employers to Discriminate against Applicants of Protected Groups},
      url = {https://incidentdatabase.ai/cite/489},
      year = {2019},
      urldate = {November 20, 2025},
      note = {Retrieved November 2025 from 
        \url{https://incidentdatabase.ai/cite/489}}
}

@article {
      aiid:534,
      author = {Anonymous},
      editor = {Lam, Khoa},
      journal = {AI Incident Database},
      publisher = {Responsible AI Collaborative},
      title = {Incident Number 534: Facebook Alleged in Lawsuit Misleading Public about Effects of Algorithms on Children},
      url = {https://incidentdatabase.ai/cite/534},
      year = {2021},
      urldate = {January 6, 2026},
      note = {Retrieved January 2026 from 
        \url{https://incidentdatabase.ai/cite/534}}
}

@article {
      aiid:592:misID,
      author = {Lam, Khoa},
      editor = {Atherton, Daniel},
      journal = {AI Incident Database},
      publisher = {Responsible AI Collaborative},
      title = {Incident Number 592: Facial Recognition Misidentifies Pregnant Woman Leading to False Arrest in Detroit},
      url = {https://incidentdatabase.ai/cite/592},
      year = {2023},
      urldate = {December 19, 2025},
      note = {Retrieved December 2025 from 
        \url{https://incidentdatabase.ai/cite/592}}
}

@article {
      aiid:597,
      author = {Atherton, Daniel},
      editor = {Atherton, Daniel},
      journal = {AI Incident Database},
      publisher = {Responsible AI Collaborative},
      title = {Incident Number 597: Female Students at Westfield High School in New Jersey Reportedly Targeted with Deepfake Nudes},
      url = {https://incidentdatabase.ai/cite/597},
      year = {2023},
      urldate = {December 19, 2025},
      note = {Retrieved December 2025 from 
        \url{https://incidentdatabase.ai/cite/597}}
}

@article {
      aiid:608,
      author = {Atherton, Daniel},
      editor = {Atherton, Daniel},
      journal = {AI Incident Database},
      publisher = {Responsible AI Collaborative},
      title = {Incident Number 608: UnitedHealth Accused of Deploying  Allegedly Flawed AI to Deny Medical Coverage},
      url = {https://incidentdatabase.ai/cite/608},
      year = {2023},
      urldate = {December 19, 2025},
      note = {Retrieved December 2025 from 
        \url{https://incidentdatabase.ai/cite/608}}
}

@article {
      aiid:639,
      author = {Atherton, Daniel},
      editor = {Atherton, Daniel},
      journal = {AI Incident Database},
      publisher = {Responsible AI Collaborative},
      title = {Incident Number 639: Air Canada Chatbot Reportedly Provides Inaccurate Bereavement Fare Information, Leading to Customer Overpayment},
      url = {https://incidentdatabase.ai/cite/639},
      year = {2022},
      urldate = {December 2, 2025},
      note = {Retrieved December 2025 from 
        \url{https://incidentdatabase.ai/cite/639}}
}

@article {
      aiid:726,
      author = {Atherton, Daniel},
      editor = {Atherton, Daniel},
      journal = {AI Incident Database},
      publisher = {Responsible AI Collaborative},
      title = {Incident Number 726: A Self-Driving Cruise Robot Taxi Reportedly Struck and Dragged a Pedestrian 20 Feet},
      url = {https://incidentdatabase.ai/cite/726},
      year = {2023},
      urldate = {December 3, 2025},
      note = {Retrieved December 2025 from 
        \url{https://incidentdatabase.ai/cite/726}}
}

@article {
      aiid:926:MeloniDeepfake,
      author = {Atherton, Daniel},
      editor = {Atherton, Daniel},
      journal = {AI Incident Database},
      publisher = {Responsible AI Collaborative},
      title = {Incident Number 926: Giorgia Meloni Reportedly Targeted by Deepfake Pornography},
      url = {https://incidentdatabase.ai/cite/926},
      year = {2020},
      urldate = {December 19, 2025},
      note = {Retrieved December 2025 from 
        \url{https://incidentdatabase.ai/cite/926}}
}

@misc{notmyai_plataforma_intervencion_social_2021,
  author       = {{Not My A.I.}},
  title        = {Case Study: Plataforma Tecnológica de Intervención Social / Projeto Horus – Argentina and Brazil},
  year         = {2021},
  howpublished = {\url{https://notmy.ai/news/case-study-plataforma-tecnologica-de-intervencion-social-argentina-and-brazil/}},
  urldate      = {2025-12-25}
}

@article{Velkova2019AlgorithmicRM,
  title={Algorithmic resistance: media practices and the politics of repair},
  author={Julia Velkova and Anne Kaun},
  journal={Information, Communication \& Society},
  year={2019},
  volume={24},
  pages={523 - 540},
  url={https://api.semanticscholar.org/CorpusID:203072114}
}

@book{bonini_trere_2024_algorithms_of_resistance,
  author       = {Tiziano Bonini and Emiliano Treré},
  title        = {Algorithms of Resistance: The Everyday Fight Against Platform Power},
  year         = {2024},
  publisher    = {The MIT Press},
  doi          = {10.7551/mitpress/14329.001.0001},
  isbn         = {9780262377485},
  url          = {https://doi.org/10.7551/mitpress/14329.001.0001}
}

@inproceedings{IncidentstoInsights,
author = {Richards, Isabel and Benn, Claire and Zilka, Miri},
title = {From Incidents to Insights: Patterns of Responsibility following AI Harms},
year = {2025},
isbn = {9798400721403},
publisher = {Association for Computing Machinery},
address = {New York, NY, USA},
url = {https://doi.org/10.1145/3757887.3763018},
doi = {10.1145/3757887.3763018},
booktitle = {Proceedings of the 5th ACM Conference on Equity and Access in Algorithms, Mechanisms, and Optimization},
pages = {151–169},
numpages = {19},
location = {
},
series = {EAAMO '25}
}

@article{Resistanceandrefusal,
author = {Ganesh, Maya and Moss, Emanuel},
year = {2022},
month = {02},
pages = {1329878X2210762},
title = {Resistance and refusal to algorithmic harms: Varieties of ‘knowledge projects’},
volume = {183},
journal = {Media International Australia},
doi = {10.1177/1329878X221076288}
}

@article{AlgorithmAuditing,
author = {Shen, Hong and DeVos, Alicia and Eslami, Motahhare and Holstein, Kenneth},
title = {Everyday Algorithm Auditing: Understanding the Power of Everyday Users in Surfacing Harmful Algorithmic Behaviors},
year = {2021},
issue_date = {October 2021},
publisher = {Association for Computing Machinery},
address = {New York, NY, USA},
volume = {5},
number = {CSCW2},
url = {https://doi.org/10.1145/3479577},
doi = {10.1145/3479577},
journal = {Proc. ACM Hum.-Comput. Interact.},
month = oct,
articleno = {433},
numpages = {29}
}

@inproceedings{FallAlgorithm,
author = {Johnson, Nari and Moharana, Sanika and Harrington, Christina and Andalibi, Nazanin and Heidari, Hoda and Eslami, Motahhare},
title = {The Fall of an Algorithm: Characterizing the Dynamics Toward Abandonment},
year = {2024},
isbn = {9798400704505},
publisher = {Association for Computing Machinery},
address = {New York, NY, USA},
url = {https://doi.org/10.1145/3630106.3658910},
doi = {10.1145/3630106.3658910},
booktitle = {Proceedings of the 2024 ACM Conference on Fairness, Accountability, and Transparency},
pages = {337–358},
numpages = {22},
location = {Rio de Janeiro, Brazil},
series = {FAccT '24}
}

@article{ComesAfterHarm,
author = {Xiao, Sijia and Zou, Haodi and Zhang, Alice and Kumar, Deepak and Shen, Hong and Hong, Jason and Eslami, Motahhare},
year = {2025},
month = {10},
pages = {2744-2756},
title = {What Comes After Harm? Mapping Reparative Actions in AI Through Justice Frameworks},
volume = {8},
journal = {Proceedings of the AAAI/ACM Conference on AI, Ethics, and Society},
doi = {10.1609/aies.v8i3.36754}
}

@inproceedings{HCITactics,
author = {Whitney, Cedric Deslandes and Naval, Teresa and Quepons, Elizabeth and Singh, Simrandeep and Rick, Steven R and Irani, Lilly},
title = {HCI Tactics for Politics from Below: Meeting the Challenges of Smart Cities},
year = {2021},
isbn = {9781450380966},
publisher = {Association for Computing Machinery},
address = {New York, NY, USA},
url = {https://doi.org/10.1145/3411764.3445314},
doi = {10.1145/3411764.3445314},
booktitle = {Proceedings of the 2021 CHI Conference on Human Factors in Computing Systems},
articleno = {297},
numpages = {15},
location = {Yokohama, Japan},
series = {CHI '21}
}

@article{BovensAccountability,
author = {Bovens, Mark},
year = {2007},
month = {07},
pages = {447 - 468},
title = {Analysing and Assessing Accountability: A Conceptual Framework},
volume = {13},
journal = {European Law Journal},
doi = {10.1111/j.1468-0386.2007.00378.x}
}

@article{whatisAccountability,
author = {Novelli, Claudio and Taddeo, Mariarosaria and Floridi, Luciano},
year = {2023},
month = {02},
pages = {1-12},
title = {Accountability in artificial intelligence: what it is and how it works},
volume = {39},
journal = {AI \& SOCIETY},
doi = {10.1007/s00146-023-01635-y}
}

@article{Socialaccountability,
author = {Camargo, Claudia and Jacobs, Eelco},
year = {2013},
month = {02},
pages = {1-24},
title = {Working Paper 16: Social accountability and its conceptual challenges: An analytical framework},
journal = {Basel Institute on Governance Working Papers},
doi = {10.12685/bigwp.2013.16.1-24}
}

@article{bingham2023,
  author       = {Bingham, J., Andrea},
  title        = {From Data Management to Actionable Findings: A Five-Phase Process of Qualitative Data Analysis},
  journaltitle = {International Journal of Qualitative Methods},
  year         = {2023},
  volume       = {22},
  doi          = {10.1177/16094069231183620}
}

@article{Buxbaum_2025, title={Adhesive Forum Selection Agreements and Access to Justice: The Function and Limits of Anti-Waiver Protections}, volume={26}, DOI={10.1017/glj.2025.10140}, number={5}, journal={German Law Journal}, author={Buxbaum, Hannah L.}, year={2025}, pages={876–888}}

@misc{groves2023goingpublicrolepublic,
      title={Going public: the role of public participation approaches in commercial AI labs}, 
      author={Lara Groves and Aidan Peppin and Andrew Strait and Jenny Brennan},
      year={2023},
      eprint={2306.09871},
      archivePrefix={arXiv},
      primaryClass={cs.HC},
      url={https://arxiv.org/abs/2306.09871}, 
}

@online{ainow_litigating_algorithms_2018,
  author       = {{AI Now Institute}},
  title        = {Litigating Algorithms},
  year         = {2018},
  url          = {https://ainowinstitute.org/news/litigating-algorithms-3},
  organization = {AI Now Institute},
}

@inbook{Accountabilityinacomputerizedsociety,
author = {Nissenbaum, Helen},
title = {Accountability in a computerized society},
year = {1997},
isbn = {1575860805},
publisher = {Center for the Study of Language and Information},
address = {USA},
booktitle = {Human Values and the Design of Computer Technology},
pages = {41–64},
numpages = {24}
}

@article{bottomAlgorithmic,
author = {Bloch-Wehba, Hannah},
year = {2022},
month = {01},
title = {Algorithmic Governance from the Bottom Up},
journal = {BYU Law Review},
doi = {10.2139/ssrn.4054640}
}

@inproceedings{ResponsesFromBelow,
author = {DeVrio, Alicia and Eslami, Motahhare and Holstein, Kenneth},
title = {Building, Shifting, \& Employing Power: A Taxonomy of Responses From Below to Algorithmic Harm},
year = {2024},
isbn = {9798400704505},
publisher = {Association for Computing Machinery},
address = {New York, NY, USA},
url = {https://doi.org/10.1145/3630106.3658958},
doi = {10.1145/3630106.3658958},
booktitle = {Proceedings of the 2024 ACM Conference on Fairness, Accountability, and Transparency},
pages = {1093–1106},
numpages = {14},
location = {Rio de Janeiro, Brazil},
series = {FAccT '24}
}

@online{dataharmrecord,
  author       = {Data Justice Lab},
  title        = {Data Harm Record},
  url          = {https://datajusticelab.org/project/data-harm-record/},
  note         = {Accessed: 2026-03-13}
}

@online{ajl_harms,
  author       = {Algorithmic Justice League},
  title        = {Share Your Story. Spark Change.},
  url          = {https://www.ajl.org/harms},
  note         = {Accessed: 2026-03-13},
}

@article{Saturation,
author = {Saunders, Benjamin and Sim, Julius and Kingstone, Tom and Baker, Shula and Waterfield, Jackie and Bartlam, Bernadette and Burroughs, Heather and Jinks, Clare},
year = {2018},
month = {07},
title = {Saturation in qualitative research: exploring its conceptualization and operationalization},
volume = {52},
journal = {Quality \& Quantity},
doi = {10.1007/s11135-017-0574-8}
}

@book{harding2008sciences,
  author       = {Harding, Sandra},
  title        = {Sciences from Below: Feminisms, Postcolonialities, and Modernities},
  publisher    = {Duke University Press},
  year         = {2008},
  doi          = {10.2307/j.ctv11smmtn},
  url          = {https://doi.org/10.2307/j.ctv11smmtn},
  note         = {Accessed 4 March 2026},
}

@inproceedings{StakeholderInvolvement,
author = {Kallina, Emma and Singh, Jatinder},
title = {Stakeholder Involvement for Responsible AI Development: A Process Framework},
year = {2024},
isbn = {9798400712227},
publisher = {Association for Computing Machinery},
address = {New York, NY, USA},
url = {https://doi.org/10.1145/3689904.3694698},
doi = {10.1145/3689904.3694698},
booktitle = {Proceedings of the 4th ACM Conference on Equity and Access in Algorithms, Mechanisms, and Optimization},
articleno = {1},
numpages = {14},
location = {San Luis Potosi, Mexico},
series = {EAAMO '24}
}

@inproceedings{StakeholderParticipation,
author = {Kallina, Emma and Bohn\'{e}, Thomas and Singh, Jatinder},
title = {Stakeholder Participation for Responsible AI Development: Disconnects Between Guidance and Current Practice},
year = {2025},
isbn = {9798400714825},
publisher = {Association for Computing Machinery},
address = {New York, NY, USA},
url = {https://doi.org/10.1145/3715275.3732069},
doi = {10.1145/3715275.3732069},
booktitle = {Proceedings of the 2025 ACM Conference on Fairness, Accountability, and Transparency},
pages = {1060–1079},
numpages = {20},
location = {
},
series = {FAccT '25}
}

@inproceedings{Co-DesigningChecklists,
author = {Madaio, Michael A. and Stark, Luke and Wortman Vaughan, Jennifer and Wallach, Hanna},
title = {Co-Designing Checklists to Understand Organizational Challenges and Opportunities around Fairness in AI},
year = {2020},
isbn = {9781450367080},
publisher = {Association for Computing Machinery},
address = {New York, NY, USA},
url = {https://doi.org/10.1145/3313831.3376445},
doi = {10.1145/3313831.3376445},
booktitle = {Proceedings of the 2020 CHI Conference on Human Factors in Computing Systems},
pages = {1–14},
numpages = {14},
location = {Honolulu, HI, USA},
series = {CHI '20}
}

@misc{googlepair_feedback_controls,
  author       = {{Google PAIR}},
  title        = {Feedback + Control: Design feedback and control mechanisms to improve your AI and the user experience},
  year         = {2019},
  howpublished = {\url{https://pair.withgoogle.com/guidebook/chapters/feedback-and-controls/design-ai-feedback-loops}},
}

@inproceedings{WorkaroundsAIContestation,
author = {Upadhyay, Sohini and Pruss, Dasha and DeVrio, Alicia and Gajos, Krzysztof Z. and Karusala, Naveena},
title = {"It just requires so much more creativity": Barriers and Workarounds to Gathering Information for AI Contestation},
year = {2026},
isbn = {9798400722783},
publisher = {Association for Computing Machinery},
address = {New York, NY, USA},
url = {https://doi.org/10.1145/3772318.3791923},
doi = {10.1145/3772318.3791923},
booktitle = {Proceedings of the 2026 CHI Conference on Human Factors in Computing Systems},
articleno = {30},
numpages = {18},
location = {
},
series = {CHI '26}
}

@inproceedings{Whattoaccount,
author = {Wieringa, Maranke},
title = {What to account for when accounting for algorithms: a systematic literature review on algorithmic accountability},
year = {2020},
isbn = {9781450369367},
publisher = {Association for Computing Machinery},
address = {New York, NY, USA},
url = {https://doi.org/10.1145/3351095.3372833},
doi = {10.1145/3351095.3372833},
booktitle = {Proceedings of the 2020 Conference on Fairness, Accountability, and Transparency},
pages = {1–18},
numpages = {18},
location = {Barcelona, Spain},
series = {FAT* '20}
}

@incollection{chandra2019,
  author       = {Chandra, Yanto. and Shang, Liang.},
  title        = {Inductive Coding},
  booktitle    = {Qualitative Research Using R: A Systematic Approach},
  publisher    = {Springer},
  address      = {Singapore},
  year         = {2019},
  doi          = {10.1007/978-981-13-3170-1_8}
}

@book{ObfuscationBrunton,
    author = {Brunton, Finn and Nissenbaum, Helen},
    title = {Obfuscation: A User's Guide for Privacy and Protest},
    publisher = {The MIT Press},
    year = {2015},
    month = {09},
    isbn = {9780262029735},
    doi = {10.7551/mitpress/9780262029735.001.0001},
    url = {https://doi.org/10.7551/mitpress/9780262029735.001.0001},
}

@ARTICLE{DecisionProvenance,
  author={Singh, Jatinder and Cobbe, Jennifer and Norval, Chris},
  journal={IEEE Access}, 
  title={Decision Provenance: Harnessing Data Flow for Accountable Systems}, 
  year={2019},
  volume={7},
  number={},
  pages={6562-6574},
  doi={10.1109/ACCESS.2018.2887201}}

@inproceedings{Contestability,
author = {Balayn, Agathe and Pi, Yulu and Widder, David Gray and Alfrink, Kars and Yurrita, Mireia and Upadhyay, Sohini and Karusala, Naveena and Lyons, Henrietta and Turkay, Cagatay and Tessono, Christelle and Attard-Frost, Blair and Gadiraju, Ujwal},
title = {From Stem to Stern: Contestability Along AI Value Chains},
year = {2024},
isbn = {9798400711145},
publisher = {Association for Computing Machinery},
address = {New York, NY, USA},
url = {https://doi.org/10.1145/3678884.3681831},
doi = {10.1145/3678884.3681831},
booktitle = {Companion Publication of the 2024 Conference on Computer-Supported Cooperative Work and Social Computing},
pages = {720–723},
numpages = {4},
location = {San Jose, Costa Rica},
series = {CSCW Companion '24}
}

@article{Non-Use,
author = {Baumer, Eric and Burrell, Jenna and Ames, Morgan G. and Brubaker, Jed and Dourish, Paul},
year = {2015},
month = {02},
pages = {52-56},
title = {On the Importance and Implications of Studying Technology Non-Use},
volume = {22},
journal = {interactions},
doi = {10.1145/2723667}
}

@article{smuha2021beyond,
  author  = {Smuha, Nathalie A.},
  title   = {Beyond the individual: governing AI's societal harm},
  journal = {Internet Policy Review},
  volume  = {10},
  number  = {3},
  year    = {2021},
  doi     = {10.14763/2021.3.1574},
  url     = {https://policyreview.info/articles/analysis/beyond-individual-governing-ais-societal-harm},
}

@misc{beckers2025unitedhealthAI,
  author       = {Emerson, Jakob},
  title        = {Judge denies UnitedHealth's bid to limit discovery in AI coverage denial case},
  howpublished = {\url{https://www.beckerspayer.com/payer/medicare-advantage/judge-denies-unitedhealths-bid-to-limit-discovery-in-ai-coverage-denial-case/}},
  year         = {2025},
}

@inproceedings{Reviewable,
author = {Cobbe, Jennifer and Lee, Michelle Seng Ah and Singh, Jatinder},
title = {Reviewable Automated Decision-Making: A Framework for Accountable Algorithmic Systems},
year = {2021},
isbn = {9781450383097},
publisher = {Association for Computing Machinery},
address = {New York, NY, USA},
url = {https://doi.org/10.1145/3442188.3445921},
doi = {10.1145/3442188.3445921},
booktitle = {Proceedings of the 2021 ACM Conference on Fairness, Accountability, and Transparency},
pages = {598–609},
numpages = {12},
location = {Virtual Event, Canada},
series = {FAccT '21}
}

@article{cheong_2024, title={Transparency and accountability in AI systems: safeguarding wellbeing in the age of algorithmic decision-making}, volume={6}, ISSN={2673-2726}, url={http://dx.doi.org/10.3389/fhumd.2024.1421273}, DOI={10.3389/fhumd.2024.1421273}, journal={Frontiers in Human Dynamics}, publisher={Frontiers Media SA}, author={Cheong, Ben Chester}, year={2024}}

@misc{stoplapdspying,
  author = {{Stop LAPD Spying Coalition}},
  title  = {Stop LAPD Spying Coalition},
  year   = {2025},
  url    = {https://stoplapdspying.org/},
  note   = {Accessed: 2025-12-18},
}

@article{MakingChange,
author = {Claus, Laura and Tracey, Paul},
year = {2019},
month = {08},
title = {Making Change from Behind a Mask: How Organizations Challenge Guarded Institutions by Sparking Grassroots Activism},
volume = {63},
journal = {Academy of Management Journal},
doi = {10.5465/amj.2017.0507}
}

@misc{kalluri2023surveillanceaipipeline,
      title={The Surveillance AI Pipeline}, 
      author={Pratyusha Ria Kalluri and William Agnew and Myra Cheng and Kentrell Owens and Luca Soldaini and Abeba Birhane},
      year={2023},
      eprint={2309.15084},
      archivePrefix={arXiv},
      primaryClass={cs.CV},
      url={https://arxiv.org/abs/2309.15084}, 
}

@misc{jerseyeveningpost2020alevel,
  author = {{Jersey Evening Post}},
  title  = {Johnson and Williamson forced into U-turn over A-level grades},
  year   = {2020},
  url    = {https://jerseyeveningpost.com/morenews/uknews/2020/08/17/johnson-and-williamson-forced-into-u-turn-over-a-level-grades/},
}

@misc{bbcnews2020alevels,
  author = {{BBC News}},
  title  = {A‑levels and GCSEs: Government U‑turn on exam grades to use teacher assessments},
  year   = {2020},
  url    = {https://www.bbc.co.uk/news/education-53923279},
}

@inproceedings{Understandingaccountability,
author = {Cobbe, Jennifer and Veale, Michael and Singh, Jatinder},
title = {Understanding accountability in algorithmic supply chains},
year = {2023},
isbn = {9798400701924},
publisher = {Association for Computing Machinery},
address = {New York, NY, USA},
url = {https://doi.org/10.1145/3593013.3594073},
doi = {10.1145/3593013.3594073},
booktitle = {Proceedings of the 2023 ACM Conference on Fairness, Accountability, and Transparency},
pages = {1186–1197},
numpages = {12},
location = {Chicago, IL, USA},
series = {FAccT '23}
}

@inproceedings{OutliningTraceability,
author = {Kroll, Joshua A.},
title = {Outlining Traceability: A Principle for Operationalizing Accountability in Computing Systems},
year = {2021},
isbn = {9781450383097},
publisher = {Association for Computing Machinery},
address = {New York, NY, USA},
url = {https://doi.org/10.1145/3442188.3445937},
doi = {10.1145/3442188.3445937},
booktitle = {Proceedings of the 2021 ACM Conference on Fairness, Accountability, and Transparency},
pages = {758–771},
numpages = {14},
location = {Virtual Event, Canada},
series = {FAccT '21}
}

@misc{Simonite2021WiredMaven,
  author       = {Tom Simonite},
  title        = {3 Years After the Maven Uproar, Google Cozies to the Pentagon},
  year         = {2021},
  howpublished = {\url{https://www.wired.com/story/3-years-maven-uproar-google-warms-pentagon/}},
}

@article{COBBE2021105573,
title = {Artificial intelligence as a service: Legal responsibilities, liabilities, and policy challenges},
journal = {Computer Law \& Security Review},
volume = {42},
pages = {105573},
year = {2021},
issn = {2212-473X},
doi = {https://doi.org/10.1016/j.clsr.2021.105573},
url = {https://www.sciencedirect.com/science/article/pii/S0267364921000467},
author = {Jennifer Cobbe and Jatinder Singh}
}

@inproceedings{toCourts,
author = {Metcalf, Jacob and Singh, Ranjit and Moss, Emanuel and Tafesse, Emnet and Watkins, Elizabeth Anne},
title = {Taking Algorithms to Courts: A Relational Approach to Algorithmic Accountability},
year = {2023},
isbn = {9798400701924},
publisher = {Association for Computing Machinery},
address = {New York, NY, USA},
url = {https://doi.org/10.1145/3593013.3594092},
doi = {10.1145/3593013.3594092},
booktitle = {Proceedings of the 2023 ACM Conference on Fairness, Accountability, and Transparency},
pages = {1450–1462},
numpages = {13},
location = {Chicago, IL, USA},
series = {FAccT '23}
}

@article{Pi_Proctor_2025, title={Toward empowering AI governance with redress mechanisms}, volume={1}, DOI={10.1017/cfl.2025.9}, journal={Cambridge Forum on AI: Law and Governance}, author={Pi, Yulu and Proctor, Maddie}, year={2025}, pages={e24}}

@misc{Hardy2018ARChoicesRuleBlocked,
  author       = {Hardy, Benjamin},
  title        = {ARChoices rule blocked},
  howpublished = {\url{https://arknews.org/index.php/2018/05/30/archoices-rule-blocked/}},
  note         = {Accessed: 2025-12-03},
  year         = {2018}
}

@misc{unrederss2,
  author = {{United Nations Human Rights Office of the High Commissioner}},
  title = {Access to remedy and the technology sector: Basic concepts and principles},
  year = {2023},
  url = {https://www.ohchr.org/sites/default/files/Documents/Issues/Business/B-Tech/access-to-remedy-concepts-and-principles.pdf},
  urldate = {2023-12-13}
}

@inproceedings{RecourseRepair,
author = {Hopkins, Aspen and Struckman, Isabella and Klyman, Kevin and Silbey, Susan S.},
title = {Recourse, Repair, Reparation, \& Prevention: A Stakeholder Analysis of AI Supply Chains},
year = {2025},
isbn = {9798400714825},
publisher = {Association for Computing Machinery},
address = {New York, NY, USA},
url = {https://doi.org/10.1145/3715275.3732017},
doi = {10.1145/3715275.3732017},
booktitle = {Proceedings of the 2025 ACM Conference on Fairness, Accountability, and Transparency},
pages = {209–227},
numpages = {19},
location = {
},
series = {FAccT '25}
}

@misc{IHPI2018,
  author       = {{Institute for Healthcare Policy \& Innovation, University of Michigan}},
  title        = {What happens when an algorithm cuts your health care},
  year         = {2018},
  month        = {March},
  howpublished = {\url{https://ihpi.umich.edu/news/what-happens-when-algorithm-cuts-your-health-care}},
  note         = {Accessed 2025-12-08}
}

@misc{USCCR2024,
  author       = {{U.S. Commission on Civil Rights}},
  title        = {The Civil Rights Implications of the Federal Use of Facial Recognition Technology},
  year         = {2024},
  month        = {September},
  howpublished = {\url{https://www.usccr.gov/files/2024-09/civil-rights-implications-of-frt_0.pdf}},
  note         = {Accessed 2025-12-08}
}

@article{MoralCrumpleZones,
author = {Elish, Madeleine},
year = {2019},
month = {03},
pages = {40-60},
title = {Moral Crumple Zones: Cautionary Tales in Human-Robot Interaction},
volume = {5},
journal = {Engaging Science, Technology, and Society},
doi = {10.17351/ests2019.260}
}

@inbook{Belli_Gaspar_Zingales_2024, place={Cambridge}, series={Cambridge Law Handbooks}, title={Regulating Facial Recognition in Brazil: Legal and Policy Perspectives}, booktitle={The Cambridge Handbook of Facial Recognition in the Modern State}, publisher={Cambridge University Press}, author={Belli, Luca and Gaspar, Walter Britto and Zingales, Nicolo}, editor={Matulionyte, Rita and Zalnieriute, MonikaEditors}, year={2024}, pages={228–241}, collection={Cambridge Law Handbooks}}

@article{Palantirtrust,
author = {Lena Ulbricht and Simon Egbert},
title ={In Palantir we trust? Regulation of data analysis platforms in public security},
journal = {Big Data \& Society},
volume = {11},
number = {3},
pages = {20539517241255108},
year = {2024},
doi = {10.1177/20539517241255108},
URL = {https://doi.org/10.1177/20539517241255108},
eprint = { https://doi.org/10.1177/20539517241255108}
}

@article{Bovens01092010,
author = {Mark Bovens},
title = {Two Concepts of Accountability: Accountability as a Virtue and as a Mechanism},
journal = {West European Politics},
volume = {33},
number = {5},
pages = {946--967},
year = {2010},
publisher = {Routledge},
doi = {10.1080/01402382.2010.486119},
URL = {https://doi.org/10.1080/01402382.2010.486119},
eprint = {https://doi.org/10.1080/01402382.2010.486119}
}

@article{IterativeThematicInquiry,
author = {David L. Morgan and Andreea Nica},
title ={Iterative Thematic Inquiry: A New Method for Analyzing Qualitative Data},
journal = {International Journal of Qualitative Methods},
volume = {19},
pages = {1609406920955118},
year = {2020},
doi = {10.1177/1609406920955118},
URL = {https://doi.org/10.1177/1609406920955118},
eprint = {https://doi.org/10.1177/1609406920955118}
}

@article{burrell2016,
  author = {Burrell, Jenna},
  title = {How the machine `thinks': Understanding opacity in machine learning algorithms},
  journal = {Big Data \& Society},
  volume = {3},
  number = {1},
  pages = {2053951715622512},
  year = {2016},
  publisher = {SAGE Publications},
  doi = {10.1177/2053951715622512}
}

@article{Wexler2018,
  author       = {Wexler, Rebecca},
  title        = {Life, Liberty, and Trade Secrets: Intellectual Property in the Criminal Justice System},
  journal      = {Stan.L. Rev.},
  year         = {2018},
  url          = {https://scholarship.law.columbia.edu/faculty_scholarship/4678},
  note         = {Available at SSRN; accessed 2025-12-08}
}

@article{CoglianeseBenDor2021,
  author       = {Coglianese, Cary and Ben Dor, Lavi M.},
  title        = {AI in Adjudication and Administration},
  journal      = {Brooklyn Law Review},
  year         = {2021},
  url          = {https://scholarship.law.upenn.edu/faculty_scholarship/2118},
}

@article{Valentine2019,
  author       = {Valentine, Sarah},
  title        = {Impoverished Algorithms: Misguided Governments, Flawed Technologies, and Social Control},
  journal      = {Fordham Urban Law Journal},
  year         = {2019},
  url          = {https://ir.lawnet.fordham.edu/ulj/vol46/iss2/4},
  note         = {Available at SSRN: \url{https://ssrn.com/abstract=3374005}; accessed 2025-12-09}
}

@inproceedings{DarkCompanionship,
author = {Zhang, Renwen and Li, Han and Meng, Han and Zhan, Jinyuan and Gan, Hongyuan and Lee, Yi-Chieh},
title = {The Dark Side of AI Companionship: A Taxonomy of Harmful Algorithmic Behaviors in Human-AI Relationships},
year = {2025},
isbn = {9798400713941},
publisher = {Association for Computing Machinery},
address = {New York, NY, USA},
url = {https://doi.org/10.1145/3706598.3713429},
doi = {10.1145/3706598.3713429},
booktitle = {Proceedings of the 2025 CHI Conference on Human Factors in Computing Systems},
articleno = {13},
numpages = {17},
location = {
},
series = {CHI '25}
}

@inproceedings{SociotechnicalHarms,
author = {Shelby, Renee and Rismani, Shalaleh and Henne, Kathryn and Moon, AJung and Rostamzadeh, Negar and Nicholas, Paul and Yilla-Akbari, N'Mah and Gallegos, Jess and Smart, Andrew and Garcia, Emilio and Virk, Gurleen},
title = {Sociotechnical Harms of Algorithmic Systems: Scoping a Taxonomy for Harm Reduction},
year = {2023},
isbn = {9798400702310},
publisher = {Association for Computing Machinery},
address = {New York, NY, USA},
url = {https://doi.org/10.1145/3600211.3604673},
doi = {10.1145/3600211.3604673},
booktitle = {Proceedings of the 2023 AAAI/ACM Conference on AI, Ethics, and Society},
pages = {723–741},
numpages = {19},
location = {Montr\'{e}al, QC, Canada},
series = {AIES '23}
}

@inbook{extraterritorialenforcement,
author = {Czerniawski, Michal and Svantesson, Dan},
year = {2023},
month = {12},
pages = {127-153},
title = {Challenges to the extraterritorial enforcement of data privacy law - EU case study},
isbn = {978-91-89840-02-7}
}

@article{kaminski2021,
  author = {Kaminski, Margot E. and Urban, Jennifer M.},
  title = {The Right to Contest AI},
  journal = {Columbia Law Review},
  volume = {121},
  number = {7},
  year = {2021},
  note = {U of Colorado Law Legal Studies Research Paper No. 21-30},
  url = {https://ssrn.com/abstract=3965041},
  date = {November 16, 2021}
}

@inproceedings{democraticcontrol,
author = {Bogiatzis-Gibbons, Daniel James},
title = {Beyond Individual Accountability: (Re-)Asserting Democratic Control of AI},
year = {2024},
isbn = {9798400704505},
publisher = {Association for Computing Machinery},
address = {New York, NY, USA},
url = {https://doi.org/10.1145/3630106.3658541},
doi = {10.1145/3630106.3658541},
booktitle = {Proceedings of the 2024 ACM Conference on Fairness, Accountability, and Transparency},
pages = {74–84},
numpages = {11},
location = {Rio de Janeiro, Brazil},
series = {FAccT '24}
}

@inproceedings{EmpoweringIndividuals,
author = {Pi, Yulu},
title = {Empowering Individuals in Automated Decision-Making: Explainability, Contestability and Beyond},
year = {2024},
isbn = {9798400711145},
publisher = {Association for Computing Machinery},
address = {New York, NY, USA},
url = {https://doi.org/10.1145/3678884.3682043},
doi = {10.1145/3678884.3682043},
booktitle = {Companion Publication of the 2024 Conference on Computer-Supported Cooperative Work and Social Computing},
pages = {1–4},
numpages = {4},
location = {San Jose, Costa Rica},
series = {CSCW Companion '24}
}

@misc{cifor_feminist_data_manifest_no_2019,
  author       = {Cifor, Marika and Garcia, Patricia and Cowan, T.L. and Rault, J. and Sutherland, T. and Chan, A. and Rode, J. and Hoffmann, A.L. and Salehi, N. and Nakamura, L.},
  title        = {Feminist Data Manifest-No},
  year         = {2019},
  howpublished = {\url{https://www.manifestno.com/}},
}

@article{contestcon,
author = {Lyons, Henrietta and Velloso, Eduardo and Miller, Tim},
title = {Conceptualising Contestability: Perspectives on Contesting Algorithmic Decisions},
year = {2021},
issue_date = {April 2021},
publisher = {Association for Computing Machinery},
address = {New York, NY, USA},
volume = {5},
number = {CSCW1},
url = {https://doi.org/10.1145/3449180},
doi = {10.1145/3449180},
journal = {Proc. ACM Hum.-Comput. Interact.},
month = {apr},
articleno = {106},
numpages = {25}
}

@article{fenger_simonse_dutch_surveillance_welfare_2024,
  author       = {Fenger, Menno and Simonse, Robin},
  title        = {The implosion of the Dutch surveillance welfare state},
  journal      = {Social Policy \& Administration},
  year         = {2024},
  volume       = {58},
  number       = {2},
  pages        = {264--276},
  doi          = {10.1111/spol.12998},
  url          = {https://doi.org/10.1111/spol.12998}
}

@inproceedings{AccountabilityInfrastructure,
author = {Ojewale, Victor and Steed, Ryan and Vecchione, Briana and Birhane, Abeba and Raji, Inioluwa Deborah},
title = {Towards AI Accountability Infrastructure: Gaps and Opportunities in AI Audit Tooling},
year = {2025},
isbn = {9798400713941},
publisher = {Association for Computing Machinery},
address = {New York, NY, USA},
url = {https://doi.org/10.1145/3706598.3713301},
doi = {10.1145/3706598.3713301},
booktitle = {Proceedings of the 2025 CHI Conference on Human Factors in Computing Systems},
articleno = {815},
numpages = {29},
location = {
},
series = {CHI '25}
}

@article{Tieleman_2025, title={Fairness in tension: A socio-technical analysis of an algorithm used to grade students}, volume={1}, DOI={10.1017/cfl.2025.6}, journal={Cambridge Forum on AI: Law and Governance}, author={Tieleman, Merlin}, year={2025}, pages={e19}}

@article{Negotiatingevil,
author = {Crofts, Penny and van Rijswijk, Honni},
year = {2020},
month = {02},
pages = {},
title = {Negotiating 'Evil': Google, Project Maven and the Corporate Form},
volume = {2},
journal = {Law, Technology and Humans},
doi = {10.5204/lthj.v2i1.1313}
}

@misc{noyb_exercise_your_rights_2025,
  author       = {{noyb – European Center for Digital Rights}},
  title        = {Exercise your rights!},
  howpublished = {\url{https://noyb.eu/en/exercise-your-rights}},
  year         = {2025},
  note         = {Accessed: 2025-12-25}
}

\appendix
ss\section{Table of Coded case data}

\definecolor{contestant}{HTML}{7C3AED}
\definecolor{dimension}{HTML}{36698D}
\definecolor{response}{HTML}{C2410C}
\definecolor{outcome}{HTML}{32995C}
\definecolor{timegray}{gray}{0.25}

\newcommand{\contest}[1]{\textcolor{contestant}{#1}}
\newcommand{\dimensionc}[1]{\textcolor{dimension}{#1}}
\newcommand{\responsec}[1]{\textcolor{response}{#1}}
\newcommand{\outcomec}[1]{\textcolor{outcome}{#1}}
\newcommand{\timec}[1]{\textcolor{timegray}{#1}}

\begin{table}[h]
\centering
\caption{Contestants and Outcomes Data.}
\SMALL
\renewcommand{\arraystretch}{1.2}
\setlength{\tabcolsep}{3px}
\begin{tabularx}{\textwidth}{@{} l l l l l l l l @{}}
\toprule
\textbf{ID} & \textbf{Name} & \textbf{Who Contests} & \textbf{Dimension(s)} & \textbf{Response(s)} & \textbf{Outcome(s)} & \textbf{Timespan} \\
\midrule
C1 & Arkansas' RUGS Home Care & \contest{AffInd, PubInt} & \dimensionc{SysOut, SysDes, InPrac} & \responsec{JusAlg, AvScr, DHarm} & \outcomec{Aband, StrInst, Comp, Rest} & \timec{7} \\
C2 & Dutch Childcare Benefits Scandal & \contest{AffInd, PubInt, RegOvr} & \dimensionc{SysDes, InPrac} & \responsec{JusAlg, AvScr} & \outcomec{Aband, StrInst, Comp, Rest} & \timec{8} \\
C3 & Google Instant Anti-Semitism Lawsuit & \contest{PubInt} & \dimensionc{SysOut} & \responsec{JusAlg, DenRes} & \outcomec{TechRep} & \timec{0} \\
C4 & Jordan Takaful Cash Assistance Algorithm & \contest{AffInd, PubInt} & \dimensionc{SysDes, InPrac} & \responsec{ConDel, AvScr} & \outcomec{TechRep, Ongoing} & \timec{7+} \\
C5 & KW vs. Armstrong (IDHW) & \contest{AffInd, PubInt} & \dimensionc{SysOut, SysDes, InPrac} & \responsec{AvScr, BCon, ConDel} & \outcomec{Aband, Ongoing} & \timec{14+} \\
C6 & McDonald's Algorithmic Wage Theft Case & \contest{AffInd, PCol} & \dimensionc{SysOut, InPrac} & \responsec{DenRes, AvScr} & \outcomec{Comp, Rest, StrInst} & \timec{0} \\
C7 & Missouri Medicaid Algorithm Eligibility & \contest{AffInd, PubInt} & \dimensionc{SysDes} & \responsec{} & \outcomec{TechRep} & \timec{5} \\
C8 & NZ's Equity Adjuster Waitlist & \contest{AffInd, PCol} & \dimensionc{SysDes, AppCon} & \responsec{JusAlg, DHarm} & \outcomec{Aband} & \timec{1} \\
C9 & NZ's Vulnerable Children Risk Modeling & \contest{PubInt, RegOvr} & \dimensionc{SysDes, InPrac} & \responsec{JusAlg, DHarm} & \outcomec{Aband} & \timec{2} \\
C10 & SAS's EVAAS in Houston & \contest{AffInd, PCol} & \dimensionc{InPrac, SysDes} & \responsec{AvScr, BCon} & \outcomec{Aband} & \timec{3} \\
C11 & Stanford Vaccine Distribution Protest & \contest{AffInd, PCol} & \dimensionc{SysDes, InPrac} & \responsec{JusAlg, DHarm} & \outcomec{Aband} & \timec{0} \\
C12 & Teen Pregnancy Risk Prediction in Argentina & \contest{PubInt, AffInd} & \dimensionc{SysDes, InPrac} & \responsec{JusAlg, DHarm} & \outcomec{Ongoing} & \timec{8+} \\
C13 & UK A-Level Algorithm Grading Scandal & \contest{AffInd, PubInt, PCol} & \dimensionc{AppCon, SysDes} & \responsec{JusAlg, DHarm, ConDel} & \outcomec{Aband, Rest, StrInst} & \timec{0} \\
C14 & UnitedHealth Denied Medical Coverage & \contest{AffInd, PCol, RegOvr} & \dimensionc{InPrac, SysOut} & \responsec{DenRes, AvScr} & \outcomec{StrInst, Ongoing} & \timec{3+} \\
C15 & Workday Hiring Algorithm Discrimination & \contest{AffInd, RegOvr} & \dimensionc{SysOut, InPrac} & \responsec{DenRes, AvScr} & \outcomec{Ongoing} & \timec{3+} \\
C16 & Amazon Worker Surveillance Fine & \contest{PCol, RegOvr} & \dimensionc{AppCon, InPrac} & \responsec{ConDel, DHarm} & \outcomec{StrInst, Ongoing} & \timec{6+} \\
C17 & Apple's CSAM photo-scanning feature & \contest{PubInt, PCol} & \dimensionc{SysDes, InPrac} & \responsec{JusAlg, DHarm} & \outcomec{Aband} & \timec{2} \\
C18 & Clearview AI NYPD Protester Identification & \contest{PubInt, AffInd} & \dimensionc{InPrac, AppCon} & \responsec{DenRes, DHarm, AvScr} & \outcomec{Sym/No, Ongoing} & \timec{1+} \\
C19 & Clearview AI Withdrawal from Canada & \contest{RegOvr, PubInt} & \dimensionc{InPrac, AppCon} & \responsec{BCon, AvScr} & \outcomec{Aband} & \timec{0} \\
C20 & Germany's HessenDATA & \contest{PubInt, RegOvr} & \dimensionc{AppCon, SysDes} & \responsec{BCon, JusAlg} & \outcomec{StrInst, Ongoing} & \timec{6+} \\
C21 & Hyderabad's \#BanTheScan Campaign & \contest{PubInt, AffInd} & \dimensionc{AppCon, InPrac} & \responsec{JusAlg, BCon} & \outcomec{Sym/No} & \timec{0} \\
C22 & McDonald's Biometric Drive-Thru Data & \contest{AffInd} & \dimensionc{InPrac} & \responsec{DenRes} & \outcomec{Sym/No} & \timec{2} \\
C23 & San Francisco Facial Recognition Ban & \contest{RegOvr, PubInt} & \dimensionc{AppCon, InPrac} & \responsec{JusAlg, BCon} & \outcomec{StrInst} & \timec{0} \\
C24 & São Paulo's Metro Intelligent Security System & \contest{PubInt, AffInd} & \dimensionc{AppCon, InPrac} & \responsec{BCon, JusAlg} & \outcomec{Ongoing} & \timec{4+} \\
C25 & Detroit PD Pregnant Woman MisID Arrest & \contest{AffInd, PubInt} & \dimensionc{SysOut, InPrac} & \responsec{JusAlg, DenRes} & \outcomec{StrInst} & \timec{1} \\
C26 & Air Canada Sued After Misleading Chatbot & \contest{AffInd} & \dimensionc{SysOut} & \responsec{DenRes} & \outcomec{Comp, TechRep} & \timec{1} \\
C27 & Character.ai Teen Harm & \contest{AffInd, PubInt} & \dimensionc{SysOut, SysDes} & \responsec{DenRes, DHarm} & \outcomec{StrInst, TechRep} & \timec{1} \\
C28 & Facebook Child Harm Misrepresentation & \contest{PubInt, RegOvr} & \dimensionc{SysDes, SysOut} & \responsec{DHarm, AvScr} & \outcomec{Aband, Ongoing} & \timec{5+} \\
C29 & Italian PM AI-Generated Deepfake & \contest{AffInd, RegOvr} & \dimensionc{SysOut} & \responsec{ConDel} & \outcomec{Ongoing} & \timec{2+} \\
C30 & NEDA "Tessa" Chatbot Shutdown & \contest{AffInd, PCol} & \dimensionc{SysOut, SysDes} & \responsec{DenRes, DHarm} & \outcomec{Aband} & \timec{0} \\
C31 & Scatter Lab's Lee-Luda Chatbot & \contest{AffInd, PubInt} & \dimensionc{SysOut, SysDes} & \responsec{JusAlg, DenRes} & \outcomec{Aband} & \timec{0} \\
C32 & Cruise Robot Taxi Dragged a Pedestrian 20 Feet & \contest{RegOvr, PubInt} & \dimensionc{SysDes, SysOut} & \responsec{DHarm, SC} & \outcomec{Aband, Comp, StrInst} & \timec{1} \\
C33 & Westfield-High Deepfake Incident & \contest{AffInd, PubInt} & \dimensionc{SysOut, AppCon} & \responsec{DHarm, DenRes} & \outcomec{StrInst} & \timec{2} \\
C34 & Anthropic Copyright Lawsuit (Bartz et al.) & \contest{AffInd, PubInt} & \dimensionc{InPrac, SysDes} & \responsec{DenRes, JusAlg} & \outcomec{Comp, Rest, StrInst} & \timec{2} \\
C35 & Clearview AI EU Data Misuse COMPlaint & \contest{PubInt, RegOvr} & \dimensionc{InPrac, SysDes} & \responsec{BCon, AvScr} & \outcomec{Sym/No, Ongoing} & \timec{4+} \\
C36 & Hollywood writers \& actors protest & \contest{PCol} & \dimensionc{AppCon, InPrac} & \responsec{JusAlg, AvScr} & \outcomec{StrInst, Ongoing} & \timec{3+} \\
C37 & Japanese Media vs. AI Copyright Infringement & \contest{PCol, PubInt} & \dimensionc{InPrac, SysDes} & \responsec{JusAlg, DenRes} & \outcomec{Ongoing} & \timec{2+} \\
C38 & OpenAI Sued Over Use of Personal Data & \contest{PubInt, AffInd} & \dimensionc{InPrac, SysDes} & \responsec{JusAlg, DenRes} & \outcomec{Ongoing} & \timec{3+} \\
C39 & Google Project Maven Boycott & \contest{PCol, PubInt} & \dimensionc{AppCon, InPrac} & \responsec{AvScr, JusAlg} & \outcomec{Aband, StrInst} & \timec{0} \\
C40 & Google Project Nimbus Boycott & \contest{PCol, PubInt} & \dimensionc{AppCon, InPrac} & \responsec{BCon, AvScr} & \outcomec{Ongoing} & \timec{5+} \\
C41 & Peculiar v. Diode Ventures Data Center & \contest{PubInt} & \dimensionc{EnvImp, InPrac} & \responsec{AvScr, DHarm} & \outcomec{Aband} & \timec{0} \\
C42 & Activist Hunger Strike Against Anthropic & \contest{PubInt} & \dimensionc{InPrac, SysDes} & \responsec{Sym/No} & \outcomec{Sym/No} & \timec{0} \\
C43 & Pause Giant AI Experiments: An Open Letter & \contest{PubInt, PCol} & \dimensionc{InPrac, SysDes} & \responsec{Sym/No} & \outcomec{Sym/No} & \timec{0} \\
\bottomrule
\end{tabularx}

\vspace{0.1cm}
\raggedright
\SMALL
\textbf{Abbreviation Keys:} \\

\textbf{\textcolor{contestant}{Who Contests:}}
\contest{\textit{AffInd}} Affected Individuals;
\contest{\textit{PubInt}} Public-Interest Groups;
\contest{\textit{PCol}} Professional, Labor, and Employee Collectives;
\contest{\textit{RegOvr}} Regulators and Oversight Bodies
\\

\textbf{\textcolor{dimension}{Dimensions:}}
\dimensionc{\textit{SysDes}} System-Design and Functions;
\dimensionc{\textit{SysOut}} System Output and Decisions;
\dimensionc{\textit{AppCon}} Application Context;
\dimensionc{\textit{InPrac}} Institutional Practice;
\dimensionc{\textit{EnvImp}} Infrastructure and Environmental Impacts \\

\textbf{\textcolor{response}{Responses:}}
\responsec{\textit{DenRes}} Denying Responsibility;
\responsec{\textit{DHarm}} Downplaying Harm 
\responsec{\textit{BCon}} Restricting Avenues for Contestation;
\responsec{\textit{JusAlg}} Justifying the Algorithm;
\responsec{\textit{AvScr}} Avoiding Scrutiny;
\responsec{\textit{ConDel}} Strategic Concession and Delay;
\responsec{\textit{Sym/No}} Symbolic or no Response
\\

\textbf{\textcolor{outcome}{Outcomes:}}
\outcomec{\textit{Rest}} Restitution;
\outcomec{\textit{Comp}} Compensation;
\outcomec{\textit{StrInst}} Structural or Institutional Outcomes;
\outcomec{\textit{Aband}} Abandonment;
\outcomec{\textit{TechRep}} Technical Repair;
\outcomec{\textit{Sym/No}} Symbolic or no Outcome

\textbf{\textcolor{timegray}{Timespan:}}
{ \timec{0} indicates that the case resolved in the same year it occurred in. 
Durations with "\timec{+}" indicate ongoing activities as of 2026.}

\label{tab:overalldata}
\end{table}

\section{Case Coding Schema}

\label{codingschema}

\begin{itemize}
    \item \textbf{Title (Description)}: The name and a brief description of the case.
    \item \textbf{Actor (Contester)}: The individual, group, or organization raising the critique or contesting the algorithm.
    \item \textbf{Actor (Contested)}: The organization or entity overseeing the algorithm’s deployment or development that is being critiqued.
    \item \textbf{Object (Focus of Contestation)}: The specific algorithmic system, dataset, or policy that is the subject of the dispute.
    \item \textbf{Harm (if any)}: The specific negative impacts or potential harms identified (e.g., discrimination, copyright infringement).
    \item \textbf{Sought Outcome}: What the contester aims to achieve (e.g., policy change, monetary damages, algorithm withdrawal).
    \item \textbf{Temporalities}: The timing nature of the contestation (e.g., Reactive vs. Proactive).
    \item \textbf{Form}: The method or venue of contestation (e.g., Litigation, Employee Walkout, Open Letter).
    \item \textbf{Process of Reconciliation}: The steps taken to resolve the conflict or address the grievances by involved parties.
    \item \textbf{Timespan}: The duration over which the contestation occurred (e.g., start and end years).
    \item \textbf{Outcome / Effect}: The final result of the contestation (e.g., Reform, Abandonment, Settlement).
    \item \textbf{Remarks}: Any additional context or notes relevant to the case coding.
    \item \textbf{Reference list}: Citations and sources used to document the case.
\end{itemize}

\renewcommand{\arraystretch}{1.5}

\subsection{Example Case Coding}
\begin{longtable}{p{5cm}p{10cm}}
\caption{Case Coding: San Francisco Facial Recognition Ban} \label{tab:sf_facial_recognition} \\
\hline
\textbf{Title (Description)} & \textbf{San Francisco Facial Recognition Ban} \\ \hline

Actor (Contester) & Public Interest Group, Government Entity, Activist(s), Affected Individuals \\ \hline
Actor (Contested) & San Francisco city government (government entity) \\ \hline
Object (Focus of Contestation) & Facial Recognition Software \\ \hline
Harm (if any) & Surveillance, Representational, Allocative \\ \hline
Sought Outcome & Contesters wanted to prevent government agencies from deploying facial-recognition technologies that threatened civil liberties, privacy, and marginalized communities. Their aim was to ban the technology before widespread adoption, ensure public oversight over surveillance acquisitions, and protect residents from biased policing, misidentification, and historically rooted surveillance abuses. They advocated for democratic governance of surveillance tools and community control over how public data and monitoring systems are used. \\ \hline
Temporalities & Proactive \\ \hline
Form & Advocacy, Political / Regulatory / Parliamentary inquiry \\ \hline
Process of Reconciliation & Pre-2019: San Francisco communities and civil-rights groups draw from decades of historical surveillance abuses—COINTELPRO, post-9/11 surveillance of Muslim Americans, racialized policing, to argue that new technologies like facial recognition would replicate and intensify longstanding harms. \newline
Early 2019: A coalition of 26 Bay Area organizations campaigns for an ordinance restricting city surveillance technologies. Concerns include algorithmic bias, racial misidentification, and the civil rights implications of state-run biometric monitoring. \newline
May 2019: The San Francisco Board of Supervisors votes 8–1 to ban government use of facial recognition and to require public approval for any new surveillance tools. Supporters cite inaccuracy, racial and gender bias, and the chilling effect on civil liberties. Critics argue for a moratorium instead of a ban. \newline
Week after initial vote: The measure proceeds to a second procedural vote and is formally incorporated into city law. \newline
Post-passage: The ban sparks nationwide debate, inspiring other cities to consider similar regulations and prompting broader public discourse on AI-enabled government surveillance. \\ \hline
Timespan & 2019 \\ \hline
Outcome / Effect & Reform \\ \hline
Remarks & Represents one of the earliest and most high-impact examples of local democratic governance intervening in AI surveillance. San Francisco’s ban is significant not only because of the technology involved, but because of the historical memory of surveillance abuses that motivated the coalition. The ordinance is explicitly preventive and rooted in deep political histories of racialized and politically targeted intelligence practices. It illustrates how AI contestation can be driven by structural, long-term concerns rather than immediate technical failures. The case also demonstrates how municipal governments can serve as critical sites of resistance against expanding surveillance infrastructures, and how community coalitions can shift policy to limit the power of both law enforcement and emerging technologies. \\
\end{longtable}

\begin{longtable}{p{5cm}p{10cm}}
\caption{Case Coding: Anthropic Copyright Lawsuit (Bartz et al.)} \label{tab:anthropic_copyright} \\
\hline
\textbf{Title (Description)} & \textbf{Anthropic Copyright Lawsuit (Bartz et al.)} \\ \hline

Actor (Contester) & Affected Individuals \\ \hline
Actor (Contested) & Anthropic (AI Developer) \\ \hline
Object (Focus of Contestation) & Data provenance, use of pirated books \\ \hline
Harm (if any) & Copyright, Allocative \\ \hline
Sought Outcome & Statutory damages, injunctive relief, destruction of infringing datasets, recognition of unlawful training-data sourcing \\ \hline
Temporalities & Reactive \\ \hline
Form & Litigation \\ \hline
Process of Reconciliation & 2023--2024: Authors and publishers file suit alleging Anthropic used copyrighted books (purchased \& destroyed; plus ~7M pirated copies). \newline
Jun 2025: Judge Alsup’s summary judgment: use of lawfully purchased books = fair use; use of pirated books = not fair use. Trial on pirated works scheduled. \newline
Aug 2025: Parties announce settlement in principle. \newline
Sep 5, 2025: Settlement details filed. \newline
Final Settlement Terms: Anthropic pays \$1.5B minimum (plus \$3,000 per work over 500,000 items). Only past liability released. Future outputs-infringement claims remain allowed. Anthropic must destroy datasets containing pirated materials and certify deletion. \\ \hline
Timespan & 2023--2025 \\ \hline
Outcome / Effect & Repair, Reform \\ \hline
Remarks & Largest copyright settlement in AI history. Establishes benchmark damages for pirated-data training. Reinforces separation between “lawful acquisition” vs “pirated datasets.” 

\end{longtable}

\section{Types of Contesting AI Actors}

\begin{table}[H]
\centering
\caption{Types of AI Contesting Actors}
\label{tab:contesting_actors}
\begin{tabular}{p{4cm}p{6cm}p{5cm}}
\hline
\textbf{Actor Type} & \textbf{Notes} & \textbf{Example(s)} \\
\hline
Affected Individuals Alone & Individuals directly harmed by AI who act independently to seek redress, typically without organizational support or significant resources. & Shannon Carpenter (McDonald’s Biometric Drive-Thru Data case) \\
\hline
High-Profile Individuals & Public figures whose visibility or structural power amplifies contestation, allowing them to generate disproportionate attention or institutional pressure. & Giorgia Meloni (Deepfake generation case) \\
\hline
Affected Individuals Supported by Public-Interest Groups & Individuals who collaborate with NGOs or advocacy organizations to pursue contestation, gaining resources, scale, and institutional leverage. & Arkansas RUGS Home Care case, Character.ai Teen Harm case \\
\hline
Public-Interest Groups on Behalf of the Public & Organizations that independently initiate contestation to address systemic or diffuse AI harms, representing broader public interests. & São Paulo Metro Facial Recognition case\\
\hline
Private-Sector Entities / Industry Groups & Companies, professional associations, or industry bodies that contest AI use when it threatens sector norms, professional standards, or market dynamics. & Japan Newspaper Publishers, Hollywood writers’ and actors’ unions contesting AI in entertainment \\
\hline
Public-Sector Entities & Regulatory or oversight bodies that contest AI harms through formal investigations, enforcement, or independent actions, using institutional authority. & EEOC investigations, Clearview AI regulatory scrutiny \\
\hline
\end{tabular}
\end{table}

\section{Case Collection}
\label{CaseSelection}

\onecolumn
\begin{longtable}{p{0.5cm} p{4cm} p{11cm}}

& \textbf{Case Name} & \textbf{Case Synopsis} \\
\hline
\endhead

& \textbf{Bias \& Allocation} \\

\textbf{ 1 } & Arkansas' RUGS Home Care \cite{FallAlgorithm, ArkansasMedicaidHCBSCuts} & The Arkansas Department of Human Services (DHS) was sued by legal aid groups and beneficiaries over the implementation of the RUGS automated scoring tool, which sharply cut in-home Medicaid hours. In 2016, nearly half of beneficiaries experienced dramatic care cuts. Discovery revealed the algorithm assigned insufficient hours and failed to account for specific conditions. After a federal court ruled the state violated due process and a state lawsuit invalidated the algorithm, the state replaced it with the ARIA tool in 2019. Litigation continued, and in 2023, a settlement was reached, forcing the state to pay damages and implement systemic reforms, ensuring detailed explanations for future reductions. \\

\textbf{ 2 } & Dutch Childcare Benefits Scandal (Tax Authority Algorithm) \cite{FallAlgorithm} & The Dutch Tax Authority implemented a discriminatory algorithm that treated dual-nationality as a risk factor, leading to wrongful repayment demands for childcare benefits from 2013 to 2019. A 2019 audit confirmed the discrimination, and a subsequent parliamentary inquiry in 2020-2021 led to the resignation of Prime Minister Mark Rutte's cabinet in 2021. Compensation and reform efforts have been ongoing since 2021. \\

\textbf{ 3 } & Google Instant Anti-Semitism Lawsuit \cite{aiid:75:googleinstant} & Google was sued by civil society groups over antisemitic search suggestions generated by its autocomplete algorithm. The lawsuit, brought in 2021, challenged Google's claim that suggestions re purely algorithmic and not manually edited. The parties eventually reached a mediated deal, though the specific terms remain undisclosed.  \\

\textbf{ 4 } & Jordan Takaful Cash Assistance Algorithm Issues \cite{FallAlgorithm} & The Jordanian government and World Bank faced criticism over the Takaful cash assistance program's automated ranking system. Since its launch in 2019, the system has been criticized for narrowing eligibility and misclassifying poor households due to proxy indicators like utility consumption. Human Rights Watch documented these exclusion errors from 2021 to 2023. As of 2024, calls for systemic reform continue, but algorithmic targeting remains in use.  \\

\textbf{ 5 } & KW vs. Armstrong (The Idaho Department of Health and Welfare) \cite{FallAlgorithm, IdahoACLUCase} & 
The Idaho Department of Health and Welfare was sued by the ACLU over its Medicaid budget tool. A 2012 lawsuit challenged the opacity of the budget formula. A 2017 settlement required the state to replace the tool and improve appeal rights. However, due to implementation delays, a court found the Department in civil contempt in March 2025 and appointed a Special Master to oversee compliance.
\\

\textbf{ 6 } & McDonald’s Algorithmic Wage Theft Case \cite{Swaine2014McDonaldsWageTheft} & McDonald's Corporation and franchise owners faced class action lawsuits filed in 2014 alleging that time-tracking software facilitated wage theft, including unpaid overtime and forced waiting time. The litigation coincided with national minimum wage debates and highlighted the scale of the alleged conduct. The legal process has involved seeking back pay and damages, with proceedings ongoing.  \\

\textbf{ 7 } & Missouri Medicaid Algorithm Eligibility (LOC) \cite{FallAlgorithm} &  

The Missouri Department of Health and Senior Services released a new eligibility algorithm for public review before deployment. Review by legal aid groups and a formal audit revealed systemic errors, and implementation was delayed. By October 2024, Missouri switched to the new algorithm, and legal services continued to support cases of lost benefits. Coverage losses have been less severe than early projections. \\

\textbf{ 8 } & New Zealand's Equity Adjuster Waitlist \cite{FallAlgorithm} & Health NZ faced political and public backlash in 2023 over an "Equity Adjustor Score" tool that prioritized surgeries based partly on ethnicity to address systemic inequities. Critics labeled the tool discriminatory. In August 2024, despite a review finding the tool ethically justifiable, Health NZ announced it would stop using the tool and revert to prior waitlist systems, citing that clinical need should be the primary factor. \\

\textbf{ 9 } & New Zealand's Vulnerable Children Predictive Risk Modeling \cite{FallAlgorithm} & The Ministry for Social Development's plan to use a predictive risk model for vulnerable children was halted. Proposed in 2012, the project faced immediate ethical concerns and was stopped by Minister Anne Tolley in 2014 ("not on my watch!"). Following media backlash in 2015, labeling the study as treating children like "lab rats," the project was abandoned in favor of human-coordinated teams. \\

\textbf{ 10 } & SAS's EVAAS in Houston \cite{aiid:96:houston} & The Houston Independent School District was sued by the teachers' federation in 2014 over the use of the EVAAS value-added algorithm for teacher evaluations. Teachers argued the proprietary algorithm was opaque; HISD argues EVAAS is proprietary (trade secret). A settlement was reached in October 2017, where the district agreed not to use EVAAS scores for firing decisions when teachers could not challenge them. \\

\textbf{ 11 } & Stanford Vaccine Distribution Algorithm Protest \cite{aiid:91, Chen2020} & Stanford Health Care faced protests from residents in December 2020 after an algorithm excluded frontline staff from the first wave of COVID-19 vaccines. Stanford apologized and stopped using the algorithm within days, switching to a human-guided allocation process. \\

\textbf{ 12 } & Teen Pregnancy Risk Prediction in Argentina \cite{FallAlgorithm} & The Ministry of Early Childhood Development in Salta, Argentina, and Microsoft were criticized for an AI model predicting teenage pregnancy risk. Launched in 2018, the system was condemned by civil society for profiling poor and indigenous girls. Despite academic and activist criticism arguing it represents biopolitical control, the system's status remains ongoing/pending as of recent reports. \\

\textbf{ 13 } & UK A-Level Algorithm Grading Scandal \cite{FallAlgorithm} & Ofqual and the UK Government faced nationwide outrage in August 2020 after an algorithm downgraded nearly 40\% of A-level grades, disproportionately affecting state schools. Following protests and legal threats, the government reversed its decision days later, reinstating teacher-assessed grades. The scandal led to the resignation of key officials and a parliamentary inquiry \\

\textbf{ 14 } & UnitedHealth used Flawed AI to Deny Medical Coverage \cite{aiid:608} & UnitedHealthcare faces a class-action lawsuit filed in November 2023 for using the nH Predict algorithm to wrongfully deny medical coverage. In February 2024, federal regulators clarified that AI cannot be used to deny coverage. A federal judge allowed the investigation to proceed in September 2025, denying the insurer's attempt to limit discovery.  \\

\textbf{ 15 } & Workday Hiring Algorithm Discrimination Case \cite{aiid:489} & Workday was sued for alleged discrimination in its hiring algorithms. A complaint filed in 2023 by a job applicant, supported by the EEOC, argued Workday acted as an employment agency. The court dismissed the "employment agency" theory in July 2024 but allowed the "agent" theory of liability. In 2025, the court granted preliminary certification for a nationwide collective action under the ADEA. \\

\hline

& \textbf{Surveillance \& Biometrics} \\
\textbf{ 16 } & Amazon Worker Surveillance Fine \cite{CNIL2024AmazonFine, Reuters2024AmazonAppealsFine} &  Amazon France Logistique was scrutinized by the French data protection authority (CNIL) for its "excessive" and "intrusive" employee monitoring systems. A 2023 investigation found that the company's use of scanners to track worker inactivity and productivity violated GDPR. In January 2024, CNIL fined Amazon €32 million, though Amazon rejected the findings and announced plans to appeal. \\ 

\textbf{ 17 } & Apple's CSAM photo-scanning feature \cite{Abelson2024AppleCSAM} &  Apple faced backlash from public interest groups and the public regarding its proposed on-device scanning algorithm for CSAM. In August 2021 Apple revealed plans to scan photos on users' devices before uploading to iCloud, sparking pushback over privacy concerns. In September 2021 Apple pause the rollout, and in December 2023, the company officially cancelled the system.  \\

\textbf{ 18 } & Clearview AI NYPD Protester Identification Case \cite{ NYPDTheCity2025, FallAlgorithm}& The NYPD and FDNY faced scrutiny for bypassing internal bans on facial recognition to identify a protester. Despite a 2020 ban on Clearview AI, an investigation revealed that in June 2024, an FDNY fire marshal secretly ran a Clearview search for the NYPD to identify a suspect from a Columbia University protest. Following an arrest in 2025 using this data, a judge dismissed the case in June 2025, citing surveillance misconduct. In July 2025, civil liberties groups condemned the practice, and the Legal Aid Society filed a lawsuit to uncover the extent of the usage. \\

\textbf{ 19 } & Clearview AI Withdrawal from Canada \cite{Warburton2020ClearviewCanada} & Clearview AI was investigated by Canadian privacy commissioners starting in early 2020 after reports that it scraped images and was used by the RCMP. In July 2020, following the investigation, Clearview announced it would cease offering facial recognition services in Canada and suspended its contract with the RCMP. Regulators continued to investigate the RCMP's use of the technology.  \\

\textbf{ 20 } & Germany's HessenDATA \cite{FallAlgorithm} & The Hesse State Police's use of the Palantir-based HessenDATA system was challenged by civil society groups. In February 2023, the Federal Constitutional Court ruled that the legal basis for the system (§ 25a HSOG) was unconstitutional and too broad. The state amended the law in late 2023, but the Society for Civil Rights (GFF) filed a new complaint in 2024, arguing the revisions still fall short of protecting rights. \\

\textbf{ 21 } & Hyderabad's \#BanTheScan Campaign \cite{FallAlgorithm} &  Law enforcement in Hyderabad was challenged by Amnesty International and the Internet Freedom Foundation over the deployment of facial recognition technology. The "BanTheScan" campaign, launched in November 2021, mapped the pervasive CCTV coverage in the city and demanded a ban on mass surveillance systems, citing risks to marginalized communities. The campaign continues to appeal for a halt to the technology's deployment absent a legal framework. \\

\textbf{ 22 } & McDonald’s Biometric Drive-Thru Data Case \cite{aiid:360:McDBiometric} & A class-action lawsuit was filed by Shannon Carpenter against McDonald's for allegedly collecting voiceprints without consent, violating BIPA. While a court allowed some claims to proceed in 2022, the parties stipulated to a dismissal in 2023 following summary judgement motions.   \\

\textbf{ 23 } & San Francisco Facial Recognition Ban \cite{Wadsworth2024SFPDFacialRecognition, Greschler2024PropESFPDFacialRecognition, Sheard2019SFPDFacialRecognition} & Following 2019 research on facial recognition risks and advocacy by the EFF, San Francisco became the first city to ban the technology in May 2019. However, a 2024 lawsuit alleged police evaded the ban by using civilians as proxies. In March 2024, voters passed Proposition E, weakening oversight by allowing SFPD to adopt surveillance tools and seek approval retroactively within a year \\

\textbf{ 24 } & São Paulo's Metro Intelligent Security System~\cite{aiid:184} & Metrô de São Paulo's deployment of a facial recognition system was challenged by consumer groups in 2022. A judge initially suspended the system, but the decision was overturned in October 2022 by a court citing public security exemptions in the data protection law (LGPD). \\

\textbf{ 25 } & Detroit PD Pregnant Woman MisID Arrest \cite{aiid:592:misID, Hill2023} &  The Detroit Police Department was sued in August 2023 following a false facial-recognition arrest of an eight-month-pregnant woman in February 2023. The police chief attributed the incident to "poor investigative work" instead of deficits in the technology. Although as federal judge dismissed major parts of the lawsuit in 2024, the case contributed to policy reforms within the Detroit PD. \\

\hline

& \textbf{Safety \& Harmful Content} \\
\textbf{ 26 } & Air Canada Sued After Chatbot Misleads Customer \cite{aiid:639} & Air Canada (AI deployer) was contested by affected individuals regarding wrong information provided by its chatbot. In 2023, Jake Moffatt filed a complaint with the British Columbia Civil Resolution Tribunal (CRT) challenging Air Canada’s refusal to honor a fare adjustment promised by the bot. Air Canada argued that the chatbot was a "separate legal entity" responsible for its own actions. However, in 2024, the tribunal directed Air Canada to compensate the customer. Following the ruling, Air Canada acknowledged the error and updated the AI system to prevent similar misrepresentations. \\

\textbf{ 27 } & Character.ai Teen Harm Case \cite{Barcott2025CharacterAI} & Character.ai is being sued by families following the suicide of 14-year-old Sewell Setzer III in February 2024, whose final conversation was with a chatbot. In October 2024, his mother filed a wrongful-death lawsuit. Additional lawsuits followed in December 2024, alleging self-harm content, alongside an investigation by the Texas Attorney General. In 2025, a U.S. Judge rejected parts of Character.ai’s motion to dismiss. The company updated its policies in July 2025 and announced in October 2025 that it would stop allowing open-ended chats for users under 18 by November 2025. \\

\textbf{ 28 } & Facebook Algorithms \& Child Harm Misrepresentation \cite{aiid:534} & Meta is facing legal action from U.S. state attorneys regarding its algorithm's impact on children. Following warnings in May 2021, Facebook paused its service "Instagram Kids". In November 2021 the Ohio Attorney General filed a lawsuit claiming Meta misled the public about algorithmic harms. The case expanded into multi-state lawsuits in 2022-23. \\

\textbf{ 29 } & Italian PM Giorgia Meloni Seeks Damages Over AI-Generated Deepfake \cite{aiid:926:MeloniDeepfake} & Two individuals were prosecuted for creating and distributing deepfake videos of Italian PM Giorgia Meloni. The videos were uploaded in 2020. In March 2024, Meloni demanded €100,000 in damages and testified in court in July 2024.  \\

\textbf{ 30 } & NEDA “Tessa” Chatbot Shutdown \cite{FallAlgorithm, Tessa2023Vice} & The National Eating Disorders Association (NEDA) faced controversy after replacing its human helpline with an AI chatbot, "Tessa," in May 2023. Activists and experts immediately found the chatbot provided harmful advice, such as weight-loss tips. NEDA suspended the chatbot in June 2023 after these failures were publicly exposed and verified, leading to the program's abandonment. \\

\textbf{ 31 } & Scatter Lab's Lee-Luda Chatbot \cite{FallAlgorithm, oh2025leeluda} & Scatter Lab faced massive backlash in late 2020 and early 2021 over its "Lee Luda" chatbot, which exhibited hate speech and used personal data from a dating app without consent. Following regulatory fines and public outrage, the chatbot was suspended in January 2021. The company later released an AI ethics checklist and a new version of the bot in 2022. \\

\textbf{ 32 } & Self-Driving Cruise Robot Taxi Dragged a Pedestrian 20 Feet \cite{aiid:726} & General Motors Cruise was scrutinized after a robotaxi dragged a pedestrian ~20 feet in October 2023. Regulators revealed Cruise had omitted this detail in its initial reports. The California DMV suspended Cruise's permit, and the company recalled its fleet to update software. In 2024, Cruise paid a \$1.5 million penalty to NHTSA and settled with the victim for \$8-12 million. \\

\textbf{ 33 } & Westfield-High Deepfake Incident \cite{IncidentstoInsights, aiid:597} & Developers of deepfake apps and users were scrutinized after AI-generated nude images of students at Westfield High School circulated in October 2023. The incident sparked parent mobilization and national reporting. In response to advocacy by victims and their families, New Jersey passed legislation in April 2025 criminalizing deceptive AI deepfakes, and the "Take It Down Act" was passed in May 2025. \\

\hline

& \textbf{Copyright \& Data Rights} \\

\textbf{ 34 } & Anthropic Copyright Lawsuit (Bartz et al.) \cite{BvA2025DAIL, BvA2025Reuters} & Anthropic faces litigation from authors and publishers over data provenance and the use of pirated books. From 2023 to 2024, plaintiffs filed suits alleging Anthropic used copyrighted books (including ~7M pirated copies) for training. In June 2025, a Judge ruled that while the use of lawfully purchased books constituted fair use, the use of pirated books did not. By August 2025, the parties announced a settlement in principle, filed in September 2025. Anthropic agreed to pay a minimum of \$1.5 billion, destroy datasets containing pirated materials, and certify their deletion, though future infringement claims regarding outputs remain allowed. \\

\textbf{ 35 } & Clearview AI EU Data Misuse Complaint \cite{Emmanouilidou2025ClearviewSolomon, NOYB2025ClearviewCriminalComplaint} & Clearview AI has been repeatedly fined by EU data protection authorities for GDPR violations regarding data provenance. Between 2022 and 2024, authorities in France, Greece, Italy, and the Netherlands issued penalties totaling ~€95.7 million. The UK ICO also issued a fine in 2022, which Clearview contested. While Clearview argues it falls outside EU/UK jurisdiction due to a lack of physical presence, Noyb filed a criminal complaint in Austria in October 2025 targeting executives, escalating the dispute into criminal law. \\

\textbf{ 36 } & Hollywood writers \& actors protest \cite{AIisSoulless} &  Unions representing writers and actors contested the unregulated use of AI by studios like Disney and Netflix. The WGA (Writers Guild of America) strike began in May 2023 and ended in September 2023 with a contract establishing AI protections. In December 2025, the UK actors union Equity began balloting members over AI scanning concerns, opening a new front in the dispute.  \\

\textbf{ 37 } & Japanese Media vs. AI Copyright Infringement \cite{japannews20250606} & Generative AI firms are being criticized by Japanese media organizations for AI outputs that provide answers without linking to original news sources. On July 17 2024, NSK issued a public statement criticizing these practices and renewed their statement on June 5 2025. The conflict is ongoing. \\

\textbf{ 38 } & OpenAI Sued Over Use of Personal Data in AI Training \cite{clarkson2023openai_privacy, Dean2023OpenAIDataLawsuit} & OpenAI was sued in June 2023 for allegedly using personal data to train its models without consent. The complaint challenges the data collection practices, while OpenAI's service terms attempt to restrict such disputes to individual arbitration. The case remains ongoing.  \\

\hline
& \textbf{Government Contracts and Infrastructure} \\

\textbf{ 39 } & Google Project Maven Boycott \cite{Google2018Maven} & Google faced an internal revolt over "Project Maven," a contract with the Pentagon to use AI for drone footage analysis. After reports revealed the project in March 2018, over 4,000 employees signed a petition demanding withdrawal, and several resigned. In June 2018, Google announced it would not renew the contract upon its 2019 expiration and subsequently published AI Principles prohibiting the design of AI for weapons. \\

\textbf{ 40 } & Google Project Nimbus Boycott (\#NoTechForApartheid) \cite{swinhoe2021nimbus} &  Google and Amazon face ongoing internal and external protests over "Project Nimbus," a \$1.2B cloud contract with the Israeli government. The contract notably includes provisions preventing the companies from denying service to specific government entities or halting services due to boycott pressure, effectively stipulating service continuity. Since its signing in May 2021, hundreds of workers have signed letters and organized demonstrations demanding withdrawal, alleging the technology aids military occupation. As of 2024, conflict between internal dissent and corporate policy remains unresolved, with companies maintaining the tool is not for military workloads. \\

\textbf{ 41 } & Peculiar v. Diode Ventures Data Center Challenge \cite{Diode2024} & Diode Ventures faced opposition in Peculiar, Missouri, regarding a proposed \$1.5B data center ("Project Harper"). In early 2024, residents mobilized after learning of the opaque project. They were assisted by a resident of Chesterton, Indiana, who had successfully opposed a similar project and shared her "No Data Center" campaign materials with the Peculiar coalition. Following this cross-state collaboration and intense public pressure, the Board of Aldermen voted in October 2024 to remove data centers from permitted land uses, effectively killing the project. \\

\hline

& \textbf{General AI Ethics Issues} \\

\textbf{ 42 } & Activist Hunger Strike Against Anthropic \cite{AnthropicHungerStrike2025} &  Anthropic was targeted by activist Guido Reichstadter in a protest regarding the race toward AGI. The activist went on a hunger strike demanding that Anthropic pause AGI development until safety and regulation structures were in place. Throughout the 30-day strike in 2025, despite media reporting, no dialogue from Anthropic was reported. \\

\textbf{ 43 } & Pause Giant AI Experiments: An Open Letter \cite{futureoflife2023pause} & AI developers were the target of an open letter published by the Future of Life Institute in March 2023, calling for a 6-month pause on training systems more powerful than GPT-4. Signed by thousands, including Elon Musk and Steve Wozniak, the letter sparked a global debate on AI risks. While it did not lead to a pause, it catalyzed discussions on regulation and safety auditing standards throughout 2023. \\

\end{longtable}

\subsection{Case Timespans}

\begin{table}[h]
\centering
\caption{Distribution of Case timespans. A duration of 0 indicates that the case resolved in the same year it occured in. Durations without a ``+'' are completed. Durations with ``+'' indicate ongoing activities as of 2026. }
\label{tab:duration_distribution}
\begin{tabular}{l c c}
\hline
\textbf{Duration (years)} & \textbf{Count} & \textbf{Case Numbers} \\
\hline
0   & 13 & 3, 6, 11, 13, 19, 21, 23, 30, 31, 39, 41, 42, 43\\
1   & 5 & 8, 25, 26, 27, 32,\\
2   & 4 & 9, 17, 22, 34\\
3   & 1 & 10 \\
5   & 1 & 7\\
7   & 1 &  1 \\
8   & 1 & 2 \\
1+  & 1 & 18 \\
2+  & 2 & 37, 39\\
3+  & 5 & 14, 15, 33, 36, 38,  \\
4+  & 2 & 24, 35 \\
5+  & 2 & 28, 40\\
6+  & 2 & 16, 20 \\ 
7+  & 1 & 4 \\
8+  & 1 & 12\\
14+ & 1 & 5 \\
\hline
\end{tabular}
\end{table}

\section{Strategies Used by Contested Actors in Response to AI Contestation}
\begin{table}
\centering
\caption{Strategies Used by Contested Actors in Response to AI Contestation} 
\label{tab:responses_from_above_hierarchical}

\begin{tabular}{>{\raggedright\arraybackslash}p{4.5cm}
                >{\raggedright\arraybackslash}p{5cm}
                >{\raggedright\arraybackslash}p{6cm}}
\hline

\textbf{Main Strategy} & \textbf{Sub-Tactic / Definition} & \textbf{Description / Examples} \\
\hline

\multirow{2}{*}{\shortstack[l]{Denying Responsibility \\ ``It's Not Our Fault''}}

& Shifting responsibility to other actors 
& Workday argued it is not an employment agency, claiming anti-discrimination laws do not apply.\\

& Blaming the AI system 
& Air Canada claimed its chatbot was a separate legal entity responsible for its own behaviour.\\
\hline

\multirow{2}{*}{\shortstack[l]{Denying or Downplaying Harm \\ ``Nothing serious happened''}}

& Direct denial
& NEDA Tessa chatbot case, where leadership accused users of lying.\\

& Downplaying the severity or scope of harm
& Westfield High School deepfake incident, where officials framed the harm as resolved, although it was not.\\

\hline

\multirow{2}{*}{\shortstack[l]{Blocking Contestations \\ ``You Can’t Challenge Us''}}
& Legal exemptions
& The state of Hesse relied on the Hessian Act to justify the deployment of HessenDATA. \\

& Contractual protections
& Project Nimbus contract ensured service continuity and protection from external pressures. \\

& Administrative/procedural defenses
& Arkansas RUGS case, where the state agency argued that plaintiffs should exhaust all administrative remedies before seeking court review. \\

\hline

\multirow{2}{*}{\shortstack[l]{Justifying the Algorithm \\ “AI is the right choice”}}
& Emphasizing benefits and positive outcomes 
& 2020 UK A-level grading controversy, where the system was described as robust and dependable. \\
& Blaming human operators to defend the system
& Detroit Police Chief attributed a wrongful arrest to poor investigative work.\\

\hline

Avoiding Scrutiny Through Intentional Opacity \newline “You Are Not Allowed to Know”
& Using proprietary or trade-secret protections 
& Houston Independent School District teacher-evaluation case, where educators were obstructed from reviewing the EVAAS value-added model due to proprietary secrecy.\\

& Procedural or institutional opacity
& Peculiar v. Diode Ventures case, where NDAs prevented sharing details. \\
\hline

Strategic Concession and Delay \newline  “We’ll fix it later (maybe)”
& Providing case-by-case exceptions while resisting system-wide audits 
& Arkansas RUGS case \\
& Postponing implementation of reforms 
& K.W. v. Armstrong \\
& Appealing regulatory fines 
& Amazon Worker Surveillance Fine \\ 
& Making only minor adjustments 
& Jordan Takaful Cash Assistance Algorithm \\
\hline

\end{tabular}

\end{table}

\section{Levels of outcomes from contestation of AI systems}
\begin{table}[h!]
\centering
\caption{Levels of outcomes from contestation of AI systems}
\label{tab:ai_contestation_outcomes}
\begin{tabular}{p{5cm} p{5cm} p{6cm}}
\hline
\textbf{Outcome Level} & \textbf{Subcategory} & \textbf{Description / Examples} \\
\hline
\multirow{2}{*}{Outcomes for Affected Individuals} 
& Restitution & Restoring the prior state or reversing a harmful decision (e.g., reinstated benefits, corrected grades) \\
& Compensation & Providing financial or material remedies to affected individuals or groups \\
\hline
\multirow{2}{*}{Outcomes for the AI System} 
& Technical Repair / Modification & The AI system is revised, redesigned, or patched to address identified harms or failures \\
& System Abandonment / Withdrawal & The system or the contested feature is suspended, discontinued, or permanently removed \\
\hline
\multirow{2}{*}{Structural or Institutional Outcomes} 
& Policy Reform & Changes to organizational, regulatory, or governance policies (e.g., new regulatory rules, legal precedent) \\
& Service or Operational Reform & Changes to how services using the AI system are delivered, managed, or structured (e.g., ceasing to serve minor users) \\
\hline
\multirow{2}{*}{Minimal or Unresolved Outcomes} 
& Symbolic or No Change & Public acknowledgments, statements, or surface-level adjustments without substantive action; or no action at all \\
& Ongoing / Pending & The case remains open, under review, or in active litigation or negotiation, with no final outcome yet \\
\hline
\end{tabular}
\end{table}

\end{document}